\documentclass[journal=jctcce,manuscript=article]{achemso}

\usepackage{enumerate}
\usepackage{amsmath}
\usepackage{mathrsfs}
\usepackage{amssymb}
\usepackage{subfig}
\usepackage{mciteplus}
\usepackage{natbib}
\usepackage{float}
\usepackage{caption}
\usepackage{setspace}
\usepackage{color}

\author{Jerryman A. Gyamfi}
\email{jerrymanappiahene.gyamfi@kuleuven.be}

\author{Thomas-C. Jagau}
\email{thomas.jagau@kuleuven.be}
\affiliation[KUL]
{Department of Chemistry, KU Leuven, Celestijnenlaan 200F, B-3001 Leuven, Belgium}

\title[Optimization of complex absorbing potentials]{A new strategy to optimize complex absorbing 
potentials for the computation of resonance energies and widths}

\begin{document}

\begin{abstract}
Complex absorbing potentials (CAPs) are artificial potentials added to electronic Hamiltonians to make 
the wave function of metastable electronic states square-integrable. This makes electronic-structure 
theory of resonances comparable to that of bound states, thus reducing the complexity of the problem. 
However, the most often used box and Voronoi CAPs depend on several parameters that have a substantial 
impact on the numerical results. Among these parameters are the CAP strength and a set of spatial 
parameters that define the onset of the CAP. It has been a common practice to minimize the perturbation 
of the resonance states due to the CAP by optimizing the strength parameter while fixing the onset 
parameters although the performance of this approach strongly depends on the chosen onset. 

Here we introduce a more general approach that allows one to optimize not only the CAP strength but 
also the spatial parameters. We show that fixing the CAP strength and optimizing the spatial parameters 
is a reliable way for minimizing CAP perturbations. We illustrate the performance of this new approach 
by computing resonance energies and widths of the temporary anions of dinitrogen, ethylene, and formic 
acid. This is done at the Hartree-Fock and equation-of-motion coupled-cluster singles and doubles levels 
of theory, using full and projected box and Voronoi CAPs.     
\end{abstract}

\maketitle 


\section{Introduction} \label{sec:intro}

Owing to their role in diverse fields of science including biochemistry,\cite{boudaiffa00,simons07} 
astrophysics,\cite{millar17} and plasma chemistry\cite{stoffels01} research on temporary anions (TAs) 
has seen a significant increase over the last decades.\cite{herbert15,ingolfsson19} TAs are metastable 
electronic states, also called resonances,\cite{moiseyev11,jagau17,jagau22} with a lifetime in the 
range of femto- to milliseconds that are formed when a molecule with zero electron affinity captures 
an electron. Because resonances are embedded in the continuum of the electronic Hamiltonian, it is 
challenging to study TAs by means of conventional quantum chemistry methods that were designed 
for bound states.

Besides approaches based on scattering theory, \cite{bardsley68,taylor71} complex-variable techniques 
offer a way to study TAs. Here, the objective is to make the diverging resonance wave functions 
square-integrable so that electronic-structure methods similar to those developed for bound states 
can be applied. Among complex-variable techniques are complex scaling,\cite{aguilar71,balslev1971,
moiseyev98a} where the electronic coordinates in the Hamiltonian are scaled by a complex number, 
the method of complex basis functions,\cite{mccurdy78,moiseyev79,white15} where one employs 
Gaussian basis functions with complex-scaled exponents, and complex absorbing potentials\cite{
jolicard85,jolicard86,riss93,muga04} (CAPs), where an imaginary potential is added to the Hamiltonian. 
These methods all come with the advantage that explicit representation of the continuum is not 
necessary for the description of a resonance but they render the Hamiltonian non-Hermitian. In 
non-Hermitian quantum mechanics,\cite{moiseyev11} the usual inner product is replaced by the 
so-called c-product\cite{moiseyev78} and expectation values can, in general, be complex-valued. 

Among the different complex-variable techniques, CAP methods are most widely used to study TAs. 
By creating an absorbing region where artificial damping takes place, the CAP makes the diverging 
wave function of a TA square-integrable so that it can be represented in a basis set of Gaussian 
functions.\cite{riss93} The electronic Hamiltonian $H$ is modified according to
\begin{equation} \label{eq:CAP1}
H_\text{CAP} = H - i \, \eta \, W \ ,
\end{equation}
where $\eta$ is usually a real scalar and $W$ a Hermitian operator. The associated Schr\"odinger 
equation
\begin{equation} \label{eq:siegert}
H_\text{CAP} | \Psi \rangle = (E_r - i \, \Gamma/2) | \Psi \rangle
\end{equation}
has complex Siegert energies,\cite{siegert39} where $E_r$ is the resonance energy and $\Gamma$ 
is the resonance width, which is related to the lifetime $\tau$ as $\Gamma = 1/\tau$.

Different functional forms have been proposed for the CAP $W$.\cite{kosloff86,neuhauser89a,
neuhauser89b,riss93,macias94,riss95,riss96,riss98,moiseyev98b,muga04,sommerfeld15} In this work, 
we shall be concerned with box CAPs\cite{riss93} and smooth Voronoi CAPs,\cite{sommerfeld15} 
which are most widely used to study molecular TAs. For both types of CAP, $W$ is a one-electron 
operator that is further specified by a set of parameters, which define the spatial boundary between 
two regions: the outer one where the CAP is active and the inner one where it is not. With a box 
CAP, the boundary is a cuboid box, which in Cartesian coordinates is defined by three parameters 
$r^0_x, r^0_y, r^0_z$. Most often, the CAP is chosen to be quadratic in the electronic coordinates 
even though higher powers have been used as well.\cite{riss93} The operator $W$ is thus defined as 
\begin{align} \label{eq:CAP2a}
W &= \sum_{\alpha=x,y,z} W_\alpha~, \displaybreak[0]\\
W_\alpha &= \begin{cases} ( | r_\alpha - o_\alpha | - r_\alpha^0 )^2 
& \text{if} \; | r_\alpha - o_\alpha | > r_\alpha^0 \\ 0 & \text{else} \end{cases} \label{eq:CAP2b} 
\end{align}
with $r_\alpha$ as the electronic coordinate along axis $\alpha$ and the vector $(o_x, o_y, o_z)$ 
as the origin of the CAP. 

With the smooth Voronoi CAP, the boundary is constructed from each nucleus' Voronoi cell with 
the sharp edges between the cells smoothed out.\cite{sommerfeld15} Here, the operator $W$ 
depends on a single spatial parameter $r^0$ and is defined as\cite{sommerfeld15} 
\begin{equation} \label{eq:CAP3}
W(\mathbf{r}) = \begin{cases} ( r_\text{av}(\mathbf{r}) - r^0 )^2 & \text{if} \; r_\text{av} > r^0 \\ 0 & 
\text{else} \end{cases}
\end{equation}
where the weighted average of electron-nuclei distances $r_\text{av}$ is given as 
\begin{equation} \label{eq:CAP4}
r_\text{av} (\mathbf{r})= \sqrt{\sum_I w_I(\mathbf{r}) \, r_I^2(\mathbf{r}) \big/ \sum_I w_I(\mathbf{r})} 
\end{equation}
and the weights $w_I$ are given as $w_I = (r_I^2 - r_\text{nearest}^2 + 1)^{-2}$ with 
$r_\text{nearest}(\mathbf{r})= \text{min}_I (| \mathbf{r} - \mathbf{R}_I |)$. Smooth Voronoi CAPs 
are especially recommended for molecules whose structure is incompatible with a cuboid box. 

The introduction of the CAP can lead to a strong perturbation of the physical Hamiltonian $H$ from Eq. 
\eqref{eq:CAP1} and, consequently, a wrong description of the resonance. For example, in our recent 
work on molecular dynamics of TAs,\cite{gyamfi22} we showed how using unsuitable CAP parameters 
may lead to an incorrect description of the time evolution. To obviate such problems, it is imperative 
to make sure that the CAP is just a small perturbation to $H$. How to achieve this has been an important 
subject since the advent of the CAP method.\cite{riss93,macias94,riss96,riss98,zhou12,jagau14}

In molecular electronic-structure calculations, the prevailing way of minimizing the perturbation is 
through optimizing only the strength parameter $\eta$, which is done according to\cite{riss93} 
\begin{equation} \label{eq:etaopt}
\text{min} \, | \eta \, dE(\eta)/d\eta |~.
\end{equation}
In contrast, there is no consensus on how to choose the onset parameters $\mathbf{r}^0$. For 
box CAP calculations on TAs, it has been suggested to use the second moment of the electron 
density of the parent neutral molecule in order to avoid optimization of $\mathbf{r}^0$ on a 
case-by-case basis.\cite{jagau14,zuev14,benda17} This approach has proved useful in the study 
of electronic resonances but it is undeniable that in some cases the resonance is strongly perturbed 
by the CAP. Even worse, in other cases one fails to determine an optimal value of $\eta$ by means 
of Eq. \eqref{eq:etaopt} or one is unable to distinguish resonances and pseudo-continuum states.  

In this work, we explore an alternative to Eq. \eqref{eq:etaopt}: We use a more general criterion 
that has the expectation value of the CAP, $\langle - i \, \eta \, W \rangle$, at its center. In this way, 
it becomes possible to optimize any CAP parameter, not just $\eta$. Specifically, we show that 
optimizing the onset $\mathbf{r}^0$ while keeping $\eta$ fixed is an effective strategy for determining 
energies and widths of TAs. 

The remainder of the article is organized as follows: In Sec. \ref{sec:theory} we give a short review 
on the derivation of Eq. \eqref{eq:etaopt} and introduce our new criterion. Sec. \ref{sec:compd} 
summarizes the details of some illustrative calculations on the TAs of dinitrogen, ethylene, and 
formic acid, the results of which are presented in Sec. \ref{sec:res}. This is followed by final 
remarks in Sec. \ref{sec:conc}.


\section{Theory} \label{sec:theory}
\subsection{Optimization of the CAP strength using the established procedure} \label{sec:eta}

We start by briefly revisiting the origin of Eq. \eqref{eq:etaopt}. Our arguments below are similar 
to the original arguments put forward in Ref. \citenum{riss93}. We consider Eqs. \eqref{eq:CAP1} 
and \eqref{eq:siegert} assuming that we have fixed the spatial parameters $\mathbf{r}^0$ and 
want to minimize the CAP reflections by optimizing the value of $\eta$. We take the unknown 
true Siegert energy $E_0$ as reference, while $E(\eta)$ represents the complex-valued energy 
computed at a given $\eta$. .

We introduce an error function $\epsilon(\eta)$ as
\begin{equation} \label{eq:error}
\epsilon(\eta) := E(\eta) - E_0~, 
\end{equation}
which, naturally, is also unknown to us like $E_0$. It has been shown for a wide variety of CAPs 
including those from Eqs. \eqref{eq:CAP2a} and \eqref{eq:CAP3} that, in a complete basis set, 
$E(\eta)$ will approach $E_0$ for vanishing CAP strength,\cite{riss93} i.e., $\lim_{\eta \to 0} 
\epsilon(\eta) = 0$. For a finite basis set, we may expand $\epsilon(\eta)$ into a Maclaurin series 
because we expect $\eta_\text{opt}$ to be in the neighborhood of $\eta=0$. This yields
\begin{equation} \label{eq:expand}
\epsilon(\eta) = \epsilon(0) + \sum_{n=1} \frac{\eta^n}{n!} \, \frac{d^n\epsilon(\eta')}{d\eta^{'n}} 
\Big\vert_{\eta'=0} \ .
\end{equation}
The constant $\epsilon(0)=E(\eta=0) - E_0$ is also unknown. In first order of $\eta$, the absolute 
error $|\epsilon|$ has the expression
\begin{equation} \label{eq:triangle}
| \epsilon(\eta) | = | E(\eta) - E_0 | = \bigg| \epsilon(0) + \eta \frac{d\epsilon(\eta')}{d\eta'} 
\Big\vert_{\eta'=0} \bigg| \;\; \leq \;\; | \epsilon(0) | + \bigg| \eta \frac{d\epsilon(\eta')}{d\eta'} 
\Big\vert_{\eta'=0} \bigg\vert
\end{equation}			
where the triangle inequality is applied in the last step. Since $| \epsilon(0) |$ is unknown, we may 
minimize $|\epsilon(\eta)|$ by minimizing $\Big| \eta \, d\epsilon(\eta')/d\eta' \vert_{\eta'=0} \Big| = 
\Big| \eta \, dE(\eta')/d\eta' \vert_{\eta'=0} \Big|$. 

If we further introduce the approximation 
\begin{equation}
\eta \, \frac{dE(\eta')}{d\eta'}\Big\vert_{\eta'=0} \approx \eta \, \frac{dE(\eta')}{d\eta'}\Big\vert_{\eta'=\eta}  
\end{equation}
then we may consider $|\epsilon(\eta)|$ as minimized if we find an $\eta$ such that Eq. \eqref{eq:etaopt} 
is satisfied. It also follows that a more accurate approximation to the true Siegert energy can be 
computed as 
\begin{equation} \label{eq:depert1}
\widetilde{E}(\eta) := E(\eta) - \eta \, dE(\eta)/d\eta \approx E_0 + \epsilon(0)
\end{equation}
where $\widetilde{E}(\eta)$ has been referred to as first-order deperturbed energy.\cite{riss93,jagau14}


\subsection{Optimization of arbitrary CAP parameters using a new error function}\label{sec:xi}

We propose that the CAP onset $\mathbf{r}^0$ be amenable to optimization when minimizing CAP 
reflections. To this end, we introduce an error function $\xi$ which gauges how strong a perturbation 
the CAP contribution $-i\eta W$ is with respect to the physical electronic Hamiltonian $H$ in Eq. 
\eqref{eq:CAP1}. Suppose we have solved Eq. \eqref{eq:siegert} within a certain electronic-structure 
model and basis set and obtained a set of Siegert energies $E_n$ and eigenfunctions $| \Psi_n \rangle$. 
Our aim is now to quantify the perturbation of $E_n$ and $| \Psi_n \rangle$ by the CAP. 

From Eq. \eqref{eq:siegert} it follows that 
\begin{equation} \label{eq:decmp1}
\langle \Psi_n | H - i \eta W  - E_t | \Psi_n \rangle = \langle \Psi_n | H - E_t | \Psi_n \rangle 
- i \, \eta \langle \Psi_n | W | \Psi_n \rangle = \Delta E_n 
\end{equation}		
where $\Delta E_n = E_n - E_t$ and $E_t$ is the real-valued threshold energy, i.e., the energy 
of the parent neutral molecule in the case of a TA. Eq. \eqref{eq:decmp1} can be decomposed into
\begin{align} \label{eq:decmp2a}
\text{Re} (\Delta E_n) & = \text{Re} \langle \Psi_n | H - E_t | \Psi_n \rangle 
+ \text{Re} \langle - i \eta W \rangle_n \\
\text{Im} (\Delta E_n) & = \text{Im} \langle \Psi_n | H | \Psi_n \rangle 
+ \text{Im} \langle - i \eta W \rangle_n 
\label{eq:decmp2b}
\end{align}
where $\langle - i \eta W \rangle_n = -i \, \eta \langle \Psi_n | W | \Psi_n \rangle$ and c-normalization 
is assumed in Eqs. \eqref{eq:decmp1}--\eqref{eq:decmp2b} and in the following. 

If the CAP is a small perturbation, we expect
\begin{align} \label{eq:xi1}
&\Big| \frac{\text{Re} \langle - i \eta W \rangle_n}{\text{Re} (\Delta E_n) } \Big| \ll 1 ~, \\
&\Big| \frac{\text{Im} \langle - i \eta W \rangle_n}{\text{Im} (\Delta E_n) } \Big| \ll 1 ~.
\label{eq:xi2}
\end{align}
If Eq. \eqref{eq:xi1} is fulfilled, we can rule out that the value computed for $\text{Re} (\Delta E_n)$ 
is an artifact of the CAP and not an approximation to the resonance position. Likewise, if Eq. 
\eqref{eq:xi2} is fulfilled, the value computed for $\text{Im} (E_n)$ approximates the resonance 
half-width and is not an artifact of the CAP. Hence, $\text{Re} \langle - i \eta W\rangle_n / \text{Re} 
(\Delta E_n)$ and $\text{Im} \langle - i \eta W \rangle_n / \text{Im} (\Delta E_n)$ are functions 
quantifying perturbation due to the CAP. The closer both quantities are to zero, the more accurate 
the computed Siegert energy is expected to be. 

We may define a single error function $\xi$ from Eqs. \eqref{eq:xi1} and \eqref{eq:xi2} as
\begin{equation} \label{eq:xi3}
\xi := \sqrt{ \Big| \frac{\text{Re} \langle - i \eta W \rangle_n}{\text{Re}(\Delta E_n)} \Big|^2 
+ \Big| \frac{\text{Im} \langle - i \eta W \rangle_n}{\text{Im} (\Delta E_n)} \Big|^2}
\end{equation}
with $\text{Re} (\Delta E_n) \neq 0$ and $\text{Im} (\Delta E_n) \neq 0$. Here too, the closer $\xi$ 
is to zero, the more accurate the computed Siegert energy is expected to be. We note that, in 
most molecular electronic-structure calculations, the contribution of the real part of the expectation 
value of the CAP to $\xi$ is much smaller than that of the imaginary part, effectively reducing Eq. 
\eqref{eq:xi3} to $\xi = | \text{Im} \langle - i \eta W \rangle_n / \text{Im} (\Delta E_n) |$. 

Eq. \eqref{eq:xi3} can be used to identify optimal values for arbitrary CAP parameters $X$ according to 
\begin{equation} \label{eq:xiopt}
\text{min} \, \xi(X) ~.
\end{equation}
$\xi$ can be computed as long as $\langle -i \eta W \rangle_n$ is defined, independent of which 
CAP parameters we seek to optimize. Eq. \eqref{eq:xiopt} thus constitutes a more general 
criterion than Eq. \eqref{eq:etaopt}. Notably, optimization of $\eta$ according to Eqs. \eqref{eq:xiopt} 
or \eqref{eq:etaopt} is not equivalent but we have not encountered any cases where the two criteria 
yield very dissimilar optimal CAP strengths, resonance positions and widths. The real 
benefit of Eq. \eqref{eq:xiopt} is, however, that the onset parameters $\mathbf{r}^0$ can be optimized 
as well. 

We note that a deperturbed energy can be defined on the basis of Eq. \eqref{eq:xiopt} similar in 
spirit to Eq. \eqref{eq:depert1}:
\begin{equation} \label{eq:depert2}
\widetilde{E}_n = E_n - \left< - i \eta W\right>_n  = \langle \Psi_n | H | \Psi_n \rangle 
\end{equation}
The deperturbed Siegert energy $\widetilde{E}_n$ is the expectation value of the physical 
Hamiltonian $H$ with respect to the eigenfunction of $H_\text{CAP}$. 


\section{Computational details} \label{sec:compd}

We illustrate the usefulness of optimizing the CAP onset parameters by means of Eq. \eqref{eq:xiopt} 
by determining Siegert energies of the temporary anions of dinitrogen ($^2\Pi_g$ state), ethylene 
($^2$B$_{2g}$ state), and formic acid ($^2$A'' state). These shape resonances, especially N$_2^-$, 
have been studied several times using CAPs combined with different electronic-structure methods.\cite{
sommerfeld98,sommerfeld01,feuerbacher03,sajeev05,pal06,ghosh12,zhou12,zuev14,jagau14b,
sommerfeld15,benda17,kunitsa17,benda18,benda18b,thodika19,phung20,gayvert22a,gayvert22b,
dempwolff22} 

We note that for a box CAP, there is an infinite number of schemes how to vary the onset $\mathbf{r}^0$.
One may fix, for example, $\eta$, $r^0_x, r^0_y$ and vary $r^0_z$. Or, alternatively, fix $\eta$, $r^0_x, 
r^0_z$ and vary $r^0_y$. It is advisable, however, to choose a scheme that is compatible with the 
molecular point group. In this article, we use in most calculations a box CAP with the same 
onset in all directions, i.e., $r^0_x = r^0_y = r^0_z = r^0$ where $r^0$ is the parameter we vary. This 
makes the comparison between results obtained with box and smooth Voronoi CAPs easy. $r^0$ is 
varied in steps of 0.1 a.u. unless indicated otherwise. 

We use Hartree-Fock (HF) as well equation-of-motion electron-attachment coupled-cluster singles 
and doubles (EOM-EA-CCSD) theory,\cite{stanton93,nooijen96} which were implemented for use with 
CAPs in the Q-Chem package.\cite{zuev14,gayvert22a,qchem50} In the CAP-HF calculations, the core 
Hamiltonian guess together with the maximum-overlap method\cite{gilbert08} is used to ensure convergence 
of the self-consistent field equations to the desired temporary anion state instead of some pseudocontinuum 
state. For the CAP-EOM-EA-CCSD calculations, we use two different approaches: In the first approach, 
referred to as ``full CAP'' in the following, CAP-HF and CAP-CCSD calculations are performed on the 
neutral molecule, followed by a CAP-EOM-EA-CCSD calculation on the anion. In the other approach, 
referred to as ``projected CAP'', CAP-free HF, CCSD, and EOM-EA-CCSD calculations are performed, 
before $H_\text{CAP}$ is constructed and diagonalized in the basis of EOM-CCSD states.\cite{
sommerfeld01,gayvert22b} The latter approach holds the advantage that the computationally expensive 
CCSD and EOM-CCSD calculations need to be carried out only once, whereas this has to be done anew 
for every value of the CAP parameters in the former approach.

All computations were done using the cc-pVTZ basis set augmented by a varying number of diffuse 
s, p, and d shells on all atoms except hydrogen. The exponents of the additional shells are even-tempered 
with a spacing of 2 using the most diffuse shell of the cc-pVTZ basis with the same angular momentum 
as starting point. For all three TAs, the equilibrium structures of the corresponding neutral molecules as 
optimized at the HF/cc-pVTZ+3p or CCSD/cc-pVTZ+3p level were used; these structures are available 
from the Supporting Information. The integrals of the CAP over atomic orbitals were evaluated numerically 
on a Becke-type grid\cite{becke88} of 99 radial points and 590 Lebedev angular points per radial 
point.\cite{zuev14,gayvert22a}  


\section{Results and discussion} \label{sec:res}
\subsection{CAP-HF calculations for the anion of dinitrogen} \label{sec:n2hf}

\begin{figure*}
\includegraphics[scale=0.42]{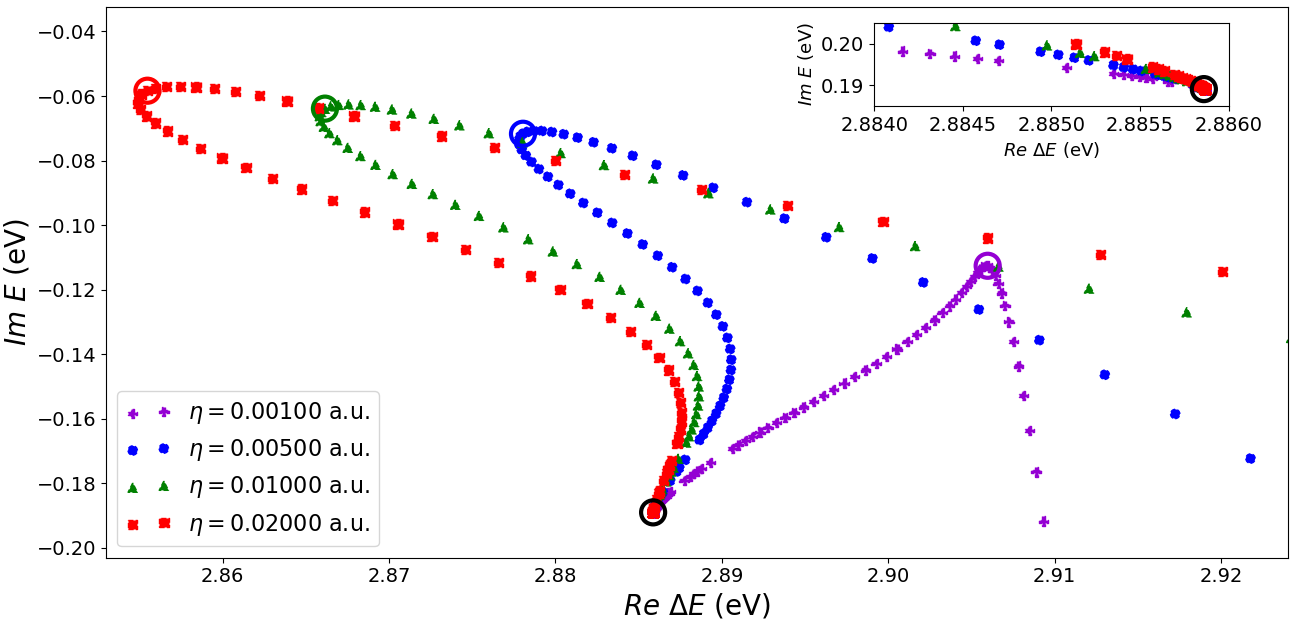}
\caption{$r^0$-Trajectories for the $^2\Pi_g$ resonance of N$_2^-$ computed with box CAPs 
of different strengths $\eta$ at the HF/cc-pVTZ+3p level of theory. Colored rings are put around 
points with minimum $\xi$, the two black rings indicate Siegert energies that are obtained over 
a wide range of large $r^0$-values for all CAP strengths. The inset shows points 
with $\text{Im}(E) > 0$ that are also obtained with large $r^0$-values.}
\label{fgr:n2hf-a} \end{figure*}

Fig. \ref{fgr:n2hf-a} shows the distribution of CAP-HF energies in the complex plane computed for 
the $^2\Pi_g$ resonance of N$_2^-$ with box CAPs with different onset parameters $r^0$. The 
different colors indicate different CAP strengths $\eta$. It is obvious that these plots are  different 
from the ones obtained by varying $\eta$ at fixed $r^0$. While $\eta$-trajectories begin at 
$\text{Im}(E)=0$ for $\eta = 0$, which corresponds to a CAP-free calculation, $r^0$-trajectories 
such as the ones shown in Fig. \ref{fgr:n2hf-a} begin at some large imaginary energy that is 
obtained for $r^0=0$ and eventually reach $\text{Im}(E)=0$, i.e., the CAP-free limit, once $r^0$ 
is so large that the CAP is invisible in the chosen basis set. 

In Fig. \ref{fgr:n2hf-a}, the points with the large spacing in the lower right correspond to low 
$r^0$-values. As we increase $r^0$, $\text{Im}(E)$ decreases until an abrupt turning point is reached, 
which corresponds to the optimal $r^0$-values according to Eq. \eqref{eq:xiopt}, i.e., a minimum 
in $\xi$. Increasing $r^0$ further does not yield the CAP-free limit $\text{Im}(E)=0$ immediately; 
instead, all $r^0$-trajectories converge to the same CAP-HF energy for all values of $r^0$ between 
ca. 8 and 20 a.u. This unphysical energy is indicated by a black circle in Fig.~\ref{fgr:n2hf-a}. Only 
with even larger $r^0$, $\text{Im}(E)=0$ is obtained. Notably, the $r^0$-trajectories obtained with 
CAP strengths of 0.02 a.u., 0.01 a.u., and 0.005 a.u. look very similar, whereas the one obtained 
with 0.001 a.u. differs significantly, suggesting that this value for $\eta$ is insufficient. 

\begin{figure*}[tbh]
\includegraphics[scale=0.42]{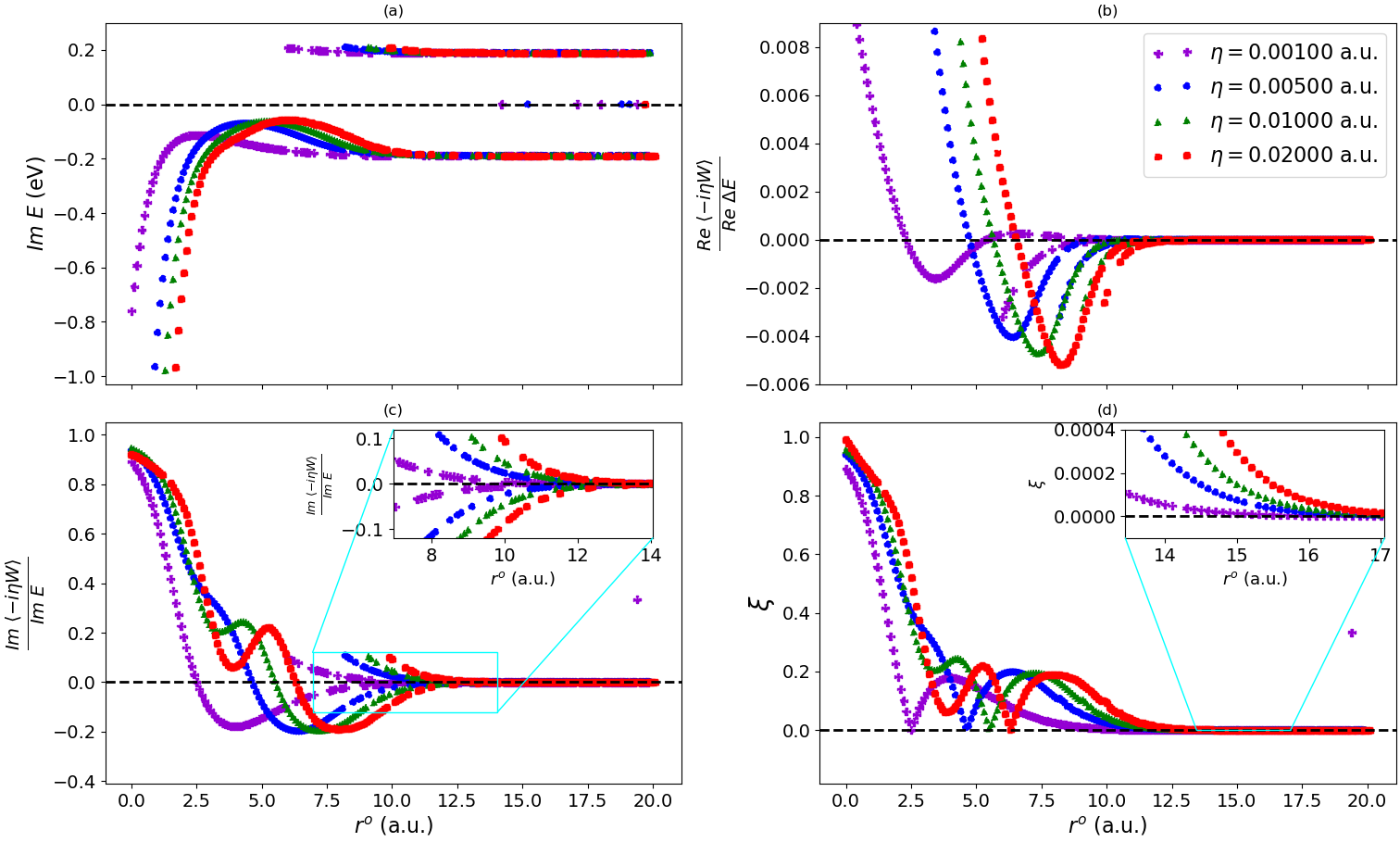}
\caption{Further analysis of the $r^0$-trajectories from Fig. \ref{fgr:n2hf-a}: (a) imaginary energy, 
(b) real part of the expectation value of the CAP, (c) imaginary part of the expectation value of 
the CAP, and (d) error function $\xi$, all as a function of $r^0$.}
\label{fgr:n2hf-b}
\end{figure*}

This is confirmed by the upper left part of Tab. \ref{tbl:n2}, which shows resonance positions 
and widths as well as the optimal $r^0$ and $\xi$ values for the CAP-HF calculations from Fig. 
\ref{fgr:n2hf-a}. If the computation with $\eta=0.001$ a.u. is neglected, the variations in $\Delta E$ 
and $\Gamma$ do not exceed 0.003 eV. By comparison, in CAP-EOM-EA-CCSD calculations 
in which $\eta$ is optimized, variations of $r^0$ by 0.5 a.u. can lead to differences of 0.05 eV 
in $\Delta E$ and $\Gamma$.\cite{zuev14} We also note that the values for the deperturbed 
resonance position and width in Tab. \ref{tbl:n2}, computed using Eq. \eqref{eq:depert2}, are 
almost identical to the original values, which is different from the approach where $\eta$ is 
optimized. Here, the first-order correction according to Eq. \eqref{eq:depert1} can make a 
significant impact. Furthermore, Tab. \ref{tbl:n2} shows that higher values for $\eta$ mandate 
a larger CAP-free region. 

Results from CAP-HF calculations with smooth Voronoi CAPs instead of box CAPs are reported 
in the second part of Tab. \ref{tbl:n2}. The computed resonance positions and widths are almost 
identical but the optimal $r^0$-values as determined according to Eq. \eqref{eq:xiopt} are 
consistently larger by about 1 a.u. as compared to the box CAP calculations. The underlying 
$r^0$-trajectories are reported in the Supporting Information and look very similar to those from 
Fig. \ref{fgr:n2hf-a}. From Tab. \ref{tbl:n2}, it is evident that CAP-HF overestimates the resonance 
position and underestimates the resonance width of N$_2^-$. This can be attributed to the neglect 
of electron correlation.\cite{jagau17} However, the CAP-HF values in Tab. \ref{tbl:n2} are in good 
agreement with CAP-HF calculations based on Eq. \eqref{eq:etaopt} that yielded 2.79 eV and 
0.13 eV for the resonance position and width using a somewhat bigger basis set.\cite{jagau17}

To analyse the $r^0$-trajectories from Fig. \ref{fgr:n2hf-a} further, we report in Fig. \ref{fgr:n2hf-b} 
several additional quantities. Panel (a) shows the imaginary energy as a function of $r^0$. This 
plot illustrates that the same CAP-HF energy is obtained between 8 a.u. and 20 a.u. for all four 
values of $\eta$ that we used. Moreover, a CAP-HF calculation run with such $r^0$-value can 
also converge to an energy with the same real part but positive imaginary part or, as a third 
solution, to the same real part and $\text{Im}(E)=0$. The latter energy is identical to the one 
obtained in the corresponding CAP-free calculation. This behavior is clearly unphysical and 
likely caused by the non-linear parameterization of the CAP-HF equations, which do not 
necessarily need to have the same solution manifold as Eq. \eqref{eq:siegert}. 

Panels (b) and (c) of Fig. \ref{fgr:n2hf-b} show the contribution of the expectation value of the 
CAP to the overall values of $\text{Re}(\Delta E)$ and $\text{Im}(E)$, respectively. As motivated 
in Sec. \ref{sec:xi}, these functions quantify the perturbation of $\text{Re}(\Delta E)$ and 
$\text{Im}(E)$ by the CAP. For all four CAP strengths we tested, there are two values of 
$r^0$ in the range 2--8 a.u. where $\text{Re} \langle i \eta W \rangle$ and $\text{Im} \langle 
i \eta W \rangle$, respectively, vanish, meaning that there is no perturbative contribution to 
$\text{Re}(\Delta E)$ and $\text{Im}(E)$. Interestingly, also the unphysical solution obtained 
at 8 a.u. $< r^0 <$ 20 a.u. is characterized by vanishing $\text{Re} \langle i \eta W \rangle$, 
$\text{Im} \langle i \eta W \rangle$, and $\xi$.

Unfortunately, the physically meaningful zeros of $\text{Re} \langle i \eta W \rangle$ and 
$\text{Im} \langle i \eta W \rangle$ occur at different values of $r^0$ so that $\xi$ computed 
using Eq. \eqref{eq:xi3} has multiple minima as displayed in panel (d) of Fig. \ref{fgr:n2hf-b}. 
This is akin to what one observes if $\eta$ is optimized according to Eq. \eqref{eq:etaopt}: 
Although $d\text{Re}(E)/d\eta$ and $d\text{Im}(E)/d\eta$ usually vanish at some $\eta$, they 
do not vanish simultaneously giving rise to multiple minima in $\eta \, dE/d\eta$. Although 
it has been argued that the evaluation of $\Delta E$ and $\Gamma$ with different CAP 
parameters may improve resonance positions and widths,\cite{jagau14} this is rarely done 
and we believe it is also not advisable in the framework of the present work. Consequently, 
we use the global minimum of $\xi$ for the values in Tab. \ref{tbl:n2} and the following tables. 


\begin{table}[tbh] \setlength{\tabcolsep}{6pt}
\caption{Vertical attachment energies $\Delta E$ and resonance widths $\Gamma$ in eV 
for the $^2\Pi_g$ resonance of N$_2^-$ computed with different CAP methods and the 
cc-pVTZ+3p basis set. $\Delta \widetilde{E}$ and $\widetilde{\Gamma}$ are the deperturbed 
vertical attachment energy and resonance width according to Eq. \eqref{eq:depert2}. Optimal 
values for $r^0$ according to Eq. \eqref{eq:xiopt} and corresponding values of $\xi$ are given 
in a.u.} \begin{small}
\begin{tabular}{lllllllllllll} \hline
$\eta$ & $r^0_\text{opt}$ & $\xi(r^0_\text{opt})$ & $\Delta E$ & $\Delta \widetilde{E}$ & $\Gamma$ & 
$\widetilde{\Gamma}$ & $r^0_\text{opt}$ & $\xi(r^0_\text{opt})$ & $\Delta E$ & $\Delta \widetilde{E}$ & 
$\Gamma$ & $\widetilde{\Gamma}$ \\ \hline
 & \multicolumn{6}{c}{box CAP/HF} & \multicolumn{6}{c}{projected box CAP/EOM-EA-CCSD, $n=2$ $^a$} \\
0.001 & 2.5 & 0.0006 & 2.906 & 2.908 & 0.225 & 0.225 & 2.600 & 0.3146 & 2.754 & 2.856 & 0.306 & 0.211 \\
0.005 & 4.6 & 0.0100 & 2.878 & 2.876 & 0.143 & 0.142 & 4.300 & 0.0243 & 2.669 & 2.703 & 0.238 & 0.233 \\
0.010 & 5.5 & 0.0064 & 2.866 & 2.864 & 0.128 & 0.127 & 5.188 & 0.0080 & 2.649 & 2.670 & 0.201 & 0.201 \\
0.020 & 6.3 & 0.0021 & 2.856 & 2.852 & 0.117 & 0.117 & 6.042 & 0.0053 & 2.633 & 2.647 & 0.173 & 0.173 \\
all & big & 0.0000 & 2.886 & 2.886 & 0.378 & 0.378 \\ \hline
 & \multicolumn{6}{c}{smooth Voronoi-CAP/HF} & \multicolumn{6}{c}{projected box 
 CAP/EOM-EA-CCSD, $n=5$ $^a$} \\
0.001 & 2.9 & 0.0102 & 2.906 & 2.907 & 0.217 & 0.215 & 2.5 & 0.4292 & 
 2.767 & 2.856 &  0.351 &  0.201 \\
0.005 & 5.7 & 0.0026 & 2.875 & 2.874 & 0.134 & 0.135 & 4.5 & 0.0129 & 
 2.667 & 2.702 &  0.296 &  0.296 \\
0.010 & 6.7 & 0.0053 & 2.862 & 2.861 & 0.116 & 0.115 &  5.3 & 0.0215 & 
 2.636 & 2.653 & 0.262 & 0.257 \\
0.020 & 7.6 & 0.0019 & 2.850 & 2.849 & 0.102 & 0.102 &  6.2 & 0.0385 & 
 2.606 &  2.621 &  0.225 &  0.233 \\
all & big & 0.0000 & 2.886 & 2.886 & 0.378 & 0.378 \\ \hline
 & \multicolumn{6}{c}{box CAP/EOM-EA-CCSD} & \multicolumn{6}{c}{ projected box 
CAP/EOM-EA-CCSD, $n=9$ $^a$} \\
0.001 & 2.4 & 0.4304 & 2.767 & 2.850 & 0.368 & 0.210 &  2.4 &  0.4327 & 
 2.767 &  2.850 &  0.365 &  0.208 \\
0.005 & 4.5 & 0.0098 & 2.672 & 2.699 & 0.306 & 0.306 &  4.5 & 0.0140 & 
 2.671 & 2.700 &  0.303 &  0.301 \\
0.010 & 5.3 & 0.0089 & 2.642 & 2.647 & 0.273 & 0.271 &  5.4 &  0.0315 & 
 2.636 & 2.654 &  0.268 &  0.276 \\
0.020 & 6.1 & 0.0080 & 2.616 & 2.613 & 0.238 & 0.239 &  6.2 &  0.0102 & 
2.607 &  2.617 &  0.237 &  0.239 \\ \hline
 & \multicolumn{6}{c}{smooth Voronoi-CAP/EOM-EA-CCSD} & \\ 
0.001 & 2.9 & 0.3664 & 2.764 & 2.847 & 0.341 & 0.216 \\
0.005 & 5.4 & 0.0059 & 2.662 & 2.677 & 0.281 & 0.281 \\
0.010 & 6.4 & 0.0174 & 2.627 & 2.630 & 0.236 & 0.240 \\
0.020 & 7.2 & 0.0346 & 2.601 & 2.595 & 0.199 & 0.192 \\ \hline 
\multicolumn{3}{c}{Reference value\cite{berman83}} & 2.32 &  & 0.41 &  \\ \hline
\end{tabular} \end{small}

\footnotesize{$^a$ Using the $n$ lowest-lying right and left EOM-EA-CCSD eigenstates as basis.}
\label{tbl:n2} \end{table}

\begin{table}[htb] \setlength{\tabcolsep}{8pt} 
\caption{Vertical attachment energies $\Delta E$ and resonance widths $\Gamma$ 
in eV for the $^2\Pi_g$ resonance of N$_2^-$ evaluated using Eq. \eqref{eq:etaopt} at fixed 
$r^0$, using Eq. \eqref{eq:xiopt} at fixed $r^0$, and using Eq. \eqref{eq:xiopt} at fixed $\eta$. 
All computations were done with box-CAP-EOM-EA-CCSD and the aug-cc-pVTZ+6s6p6d 
basis set. $\Delta \widetilde{E}$ and $\widetilde{\Gamma}$ are the deperturbed vertical 
attachment energy and resonance width according to Eqs. \eqref{eq:depert1} or 
\eqref{eq:depert2}. Values of $\eta$, $r^0$, and $\xi$ are given in a.u.}
\begin{tabular}{lllllll} \hline
$r^0$ & $\eta$ & $\xi$ & $\Delta E$ & $\Delta \widetilde{E}$ & $\Gamma$ & $\widetilde{\Gamma}$ \\ \hline
\multicolumn{7}{c}{Determined from an $\eta-$trajectory using Eq. \eqref{eq:etaopt}} \\
2.765/2.765/4.881 $^a$ & 0.00360 $^b$ & 0.1716 & 2.570 & 2.571 & 0.361 & 0.423 \\ \hline
\multicolumn{7}{c}{Determined from an $\eta-$trajectory using Eq. \eqref{eq:xiopt}} \\
2.765/2.765/4.881 $^a$ & 0.00420 $^b$ & 0.0172 & 2.571 & 2.559 & 0.356 & 0.362 \\ \hline
\multicolumn{7}{c}{Determined from an $r^0-$trajectory using Eq. \eqref{eq:xiopt}} \\
1.6/1.6/1.6 $^b$ & 0.00100 $^a$ & 0.0195 & 2.616 & 2.613 & 0.655 & 0.642 \\
3.6/3.6/3.6 $^b$ & 0.00500 $^a$ & 0.0146 & 2.571 & 2.548 & 0.319 & 0.323 \\
4.4/4.4/4.4 $^b$ & 0.01000 $^a$ & 0.0274 & 2.552 & 2.524 & 0.257 & 0.251 \\
5.3/5.3/5.3 $^b$ & 0.02000 $^a$ & 0.0161 & 2.526 & 2.502 & 0.204 & 0.207 \\ \hline
\multicolumn{3}{c}{Reference value\cite{berman83}} & 2.32 &  & 0.41 & \\ \hline
\end{tabular}

\footnotesize{$^a$ Fixed.}
\footnotesize{$^b$ Optimized.}
\label{tbl:n2b} \end{table}


\subsection{CAP-EOM-EA-CCSD calculations for the anion of dinitrogen} \label{sec:n2cc}

\begin{figure*}
\includegraphics[scale=0.42]{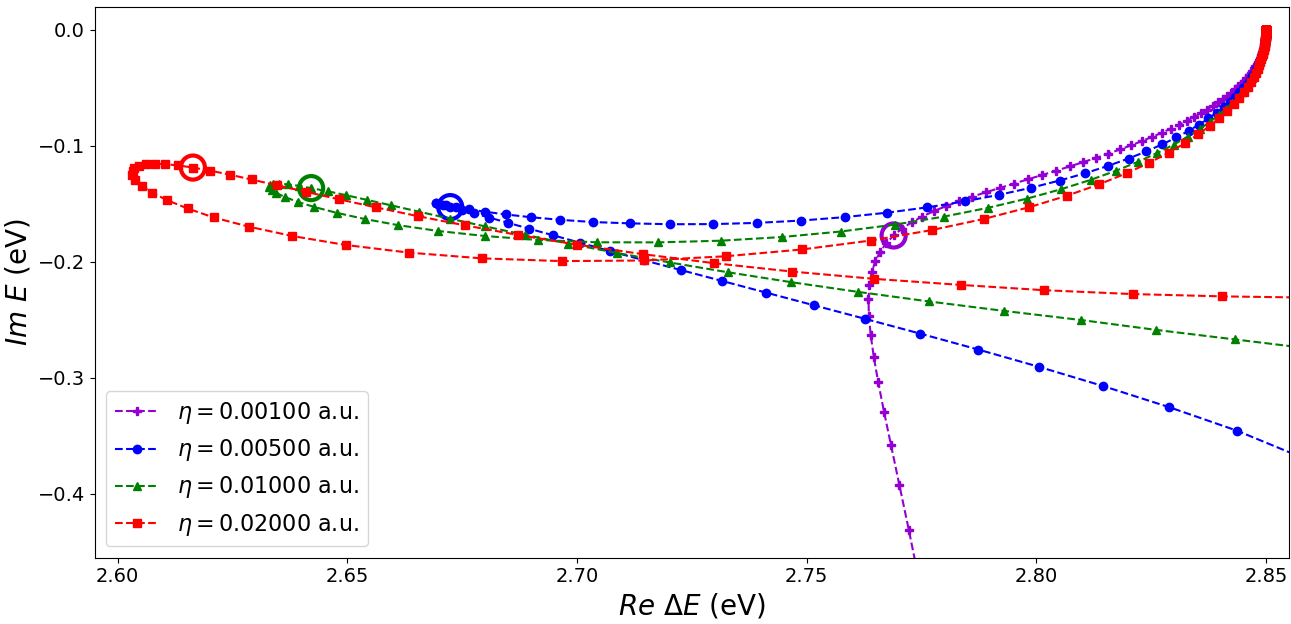}
\caption{$r^0$-Trajectories for the $^2\Pi_g$ resonance of N$_2^-$ computed with box CAPs of 
different strengths $\eta$ at the EOM-EA-CCSD/cc-pVTZ+3p level of theory. Colored rings are put 
around points with minimum $\xi$.}
\label{fgr:n2cc-a} \end{figure*}

Fig. \ref{fgr:n2cc-a} shows $r^0$-trajectories for the $^2\Pi_g$ resonance of N$_2^-$ computed 
with CAP-EOM-EA-CCSD and different CAP strengths. Resonance positions and widths from these 
calculations are reported in the third part of Tab. \ref{tbl:n2}. Similar to Fig. \ref{fgr:n2hf-a}, the data 
points in the lower right part of Fig. \ref{fgr:n2cc-a} correspond to low $r^0$-values. Upon increasing 
$r^0$, the complex energy stabilizes, eventually a turning point is reached, and $\xi$ assumes its 
minimum value. A striking difference to Fig. \ref{fgr:n2hf-a} is that the $r^0$-trajectories converge to 
the CAP-free limit ($\text{Im}(E)=0$) straight away after the turning point; we did not encounter any 
unphysical solution of the CAP-EOM-EA-CCSD equations at large $r^0$. This is no surprise 
because the CAP-EOM-CC Hamiltonian is a similarity transform of $H_\text{CAP}$ so that, in 
contrast to CAP-HF theory, additional solutions do not exist. 

\begin{figure*}
\includegraphics[scale=0.42]{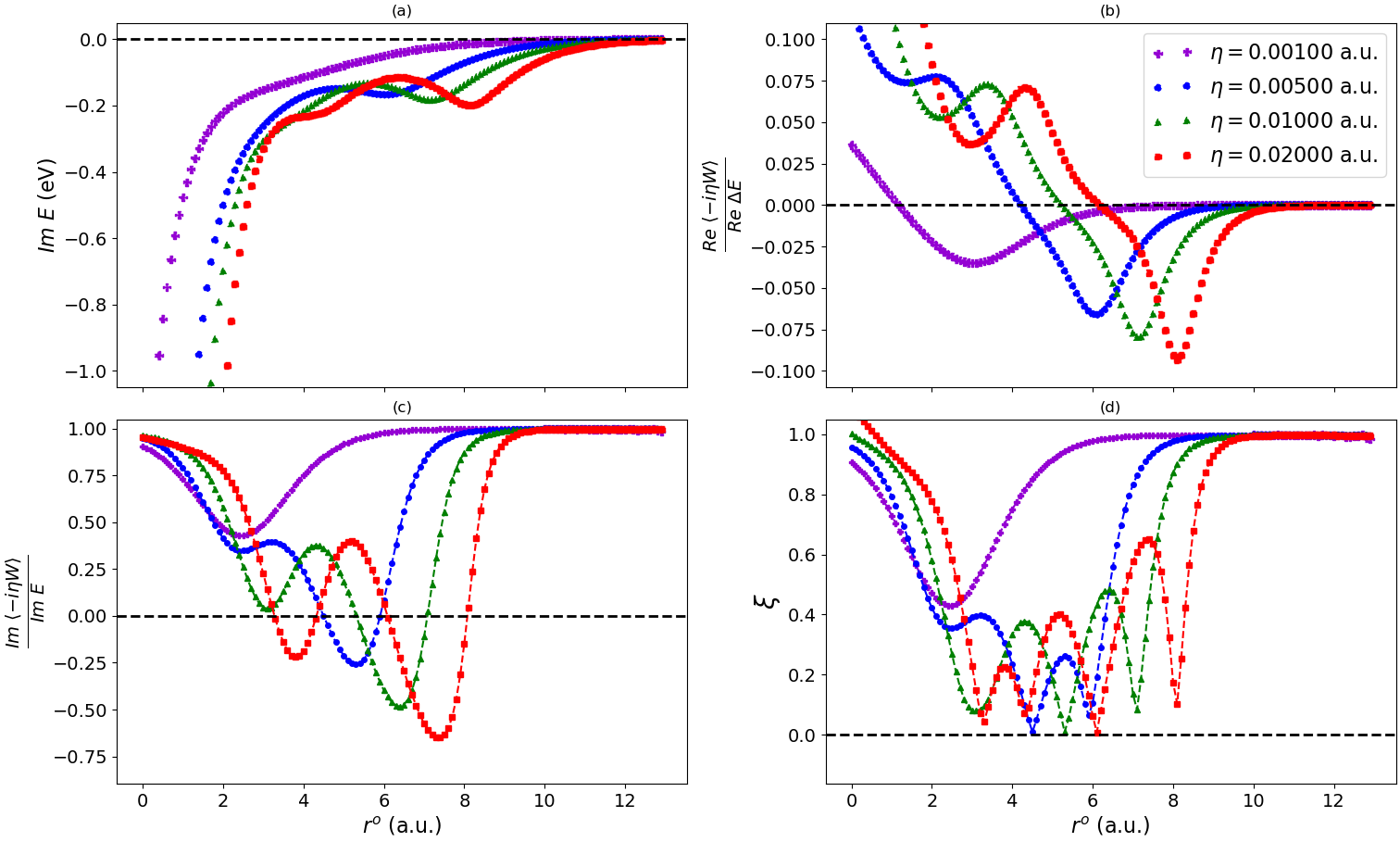}
\caption{Further analysis of the $r^0-$trajectories from Fig. \ref{fgr:n2cc-a}: (a) imaginary energy, 
(b) real part of the expecation value of the CAP, (c) imaginary part of the expectation value of the 
CAP, and (d) error function $\xi$, all as a function of $r^0$.}
 \label{fgr:n2cc-b}
\end{figure*}

The differences between CAP-EOM-EA-CCSD and CAP-HF are also evident from Fig.~\ref{fgr:n2cc-b}, 
which shows for CAP-EOM-EA-CCSD the same quantities that are shown for CAP-HF in Fig. 
\ref{fgr:n2hf-b}. The imaginary energy in panel (a) reaches the CAP-free limit between 10 and 
12 a.u. and from panel (c) it is evident that $\text{Im}(E)$ almost exclusively consists of $\text{Im} 
\langle i \eta W \rangle$ at large $r^0$-values. This implies that the deperturbed imaginary 
energy according to Eq. \eqref{eq:depert2} is practically zero, i.e., the CAP mainly acts as a 
perturbation at large $r^0$ and the resonance position and width must be evaluated at lower $r^0$. 

Interestingly, there are several $r^0$-values for which $\text{Im} \langle i \eta W \rangle$ 
vanishes in the CAP-EOM-EA-CCSD calculations, whereas this happens for exactly one 
$r^0$-value in the CAP-HF calculations reported in Fig. \ref{fgr:n2hf-b}. The corresponding 
real part of the expectation value, $\text{Re} \langle i \eta W \rangle$, which is displayed in 
panel (b) of Fig. \ref{fgr:n2cc-b}, behaves similarly for CAP-EOM-EA-CCSD and CAP-HF. 
Owing to the behavior of $\text{Im} \langle i \eta W \rangle$, $\xi$ shown in panel (d) of 
Fig.~\ref{fgr:n2cc-b} has up to four minima. 

The CAP-EOM-EA-CCSD resonance positions and widths reported in the third part of Tab. 
\ref{tbl:n2} are evaluated at the global minimum of $\xi$. These values show a more pronounced 
dependence on $\eta$ than the CAP-HF values in the upper part of the same table, while the 
optimal $r^0$ is very similar for CAP-HF and CAP-EOM-EA-CCSD. $\Delta E$ and $\Gamma$ 
evaluated with CAP strengths of 0.005 a.u. and 0.02 a.u. differ by up to 0.05 eV and 0.08 eV, 
respectively. This is akin to the sensitivity toward the CAP onset observed in CAP-EOM-EA-CCSD 
calculations in which $\eta$ is optimized\cite{zuev14} and demonstrates that choosing $\eta$ 
correctly is important in our approach where $r^0$ is optimized. By comparing the values of 
$\xi$ in Tab. \ref{tbl:n2}, we conclude that 0.02 a.u. is the best among the four CAP strengths 
we used. We further note that the correction from Eq. \eqref{eq:depert2} generally makes a 
negligible impact on positions and widths except for $\eta=0.001$ a.u. Together with the large 
value of $\xi$, this indicates that 0.001 a.u. is an insufficient CAP strength. 

The fourth part of Tab. \ref{tbl:n2} reports resonance energies and widths computed with 
CAP-EOM-EA-CCSD and a smooth Voronoi CAP instead of box CAP. The corresponding 
$r^0$-trajectories are reported in the Supporting Information and look very similar to those 
from Fig.~\ref{fgr:n2cc-a}. As we discussed for CAP-HF, optimal $r^0$-values are bigger 
by about 1 a.u. in calculations with a smooth Voronoi CAP compared to the box CAP. The 
differences in the resonance positions and widths computed with box CAP and smooth 
Voronoi CAP amount to up to 0.02 eV and are thus somewhat larger than in CAP-HF 
calculations. 

On the right hand side of Tab. \ref{tbl:n2}, we show results from projected 
CAP-EOM-EA-CCSD calculations.\cite{gayvert22b} The corresponding $r^0$-trajectories 
are available in the Supporting Information and look again similar to those from Fig. 
\ref{fgr:n2cc-a}. Using 2 CAP-free EOM-EA-CCSD eigenstates as basis, the 
resonance positions computed with the projected approach deviate by no more than 0.02 
eV from the values obtained with full CAP-EOM-EA-CCSD, but the resonance width is 
systematically underestimated by up to 0.07 eV. Using 5 EOM-EA-CCSD eigenstates as 
basis reduces the deviation in the width to less than 0.02 eV, and in a basis of 9 states 
the width is identical to values from full CAP-EOM-EA-CCSD up to 0.005 eV.

Notably, all CAP-EOM-EA-CCSD results in Table \ref{tbl:n2} differ substantially 
from the reference value, which was obtained from a model fitted to the experimental cross 
section of N$_2$-$e^-$ scattering.\cite{berman83} This can likely be attributed to the use of 
the relatively small cc-pVTZ+3p basis set. Previous CAP-EOM-EA-CCSD calculations 
demonstrated that the use of a quadruple-$\zeta$ or quintuple-$\zeta$ basis, the inclusion 
of more than three diffuse $p$-shells, and especially the inclusion of diffuse $d$-shells bring 
the position and width in better agreement with the reference value.\cite{zuev14,jagau16,gayvert22b} 

To corroborate this point and to directly compare results from Eqs. \eqref{eq:etaopt} 
and \eqref{eq:xiopt}, we carried out additional CAP-EOM-EA-CCSD calculations using an 
aug-cc-pVTZ+6s6p6d basis set, which is available from the Supporting Information. These 
results are summarized in Table \ref{tbl:n2b}. Evidently, the use of the larger basis set 
improves the resonance position significantly by ca. 0.1 eV compared to Table \ref{tbl:n2}. 
It is furthermore apparent that optimization of the CAP onset $r^0$ at fixed CAP strength 
$\eta$ and the opposite approach yield very similar resonance positions if the respective 
fixed parameter is chosen well. The resonance width is improved by the bigger basis set 
as well, but here a significant deviation of ca. 0.04 eV between optimization of $\eta$ and 
optimization of $r^0$ remains. Lastly, the results for $\Gamma$ in Table \ref{tbl:n2} also 
illustrate that Eqs. \eqref{eq:etaopt} and \eqref{eq:xiopt} are not numerically equivalent 
although the differences amount to no more than 0.005 eV.


\subsection{Anion of ethylene} \label{sec:c2h4}

\begin{table}[htb] \setlength{\tabcolsep}{6pt}
\caption{Vertical attachment energies $\Delta E$ and resonance widths $\Gamma$ in eV for 
the $^2$B$_{2g}$ resonance of C$_2$H$_4^-$ computed with different CAP methods and the 
cc-pVTZ+3p basis set. $\Delta\widetilde{E}$ and $\widetilde{\Gamma}$ are the deperturbed 
vertical attachment energy and resonance width according to Eq. \eqref{eq:depert2}. Optimal 
values for $r^0$ according to Eq. \eqref{eq:xiopt} and corresponding values of $\xi$ are given in a.u.}
\begin{small} \begin{tabular}{lllllllllllll} \hline
$\eta$ & $r^0_\text{opt}$ & $\xi(r^0_\text{opt})$ & $\Delta E$ & $\Delta\widetilde{E}$ & $\Gamma$ & 
$\widetilde{\Gamma}$ & $r^0_\text{opt}$ & $\xi(r^0_\text{opt})$ & $\Delta E$ & $\Delta\widetilde{E}$ & 
$\Gamma$ & $\widetilde{\Gamma}$ \\ \hline
 & \multicolumn{6}{c}{box CAP/HF} & \multicolumn{6}{c}{projected box CAP/EOM-EA-CCSD, $n=2$ $^a$} \\
0.001 & 2.4 & 0.1546 & 2.823 & 2.759 & 0.679 & 0.575 & 4.0 & 0.2022 & 2.189 & 2.303 & 0.340 & 0.273 \\
0.005 & 6.4 & 0.0473 & 2.674 & 2.757 & 0.550 & 0.531 & 6.0 & 0.0335 & 2.102 & 2.134 & 0.247 & 0.255 \\
0.010 & 6.9 & 0.0194 & 2.537 & 2.563 & 0.526 & 0.517 & 6.8 & 0.0362 & 2.082 & 2.098 & 0.206 & 0.199 \\
0.020 & 7.7 & 0.0073 & 2.464 & 2.475 & 0.450 & 0.448 & 7.7 & 0.0083 & 2.065 & 2.075 & 0.164 & 0.165 \\ \hline
 & \multicolumn{6}{c}{smooth Voronoi CAP/HF} & \multicolumn{6}{c}{ 
 projected box CAP/EOM-EA-CCSD, $n=5$ $^a$} \\
0.001 & 2.0 & 0.1159 & 2.821 & 2.758 & 0.641 & 0.568 &  3.9 &  0.3558 & 
2.209 &  2.309 &  0.409 &  0.264 \\
0.005 & 6.5 & 0.0196 & 2.602 & 2.653 & 0.547 & 0.547 & 6.1 & 0.0292 & 
 2.095 &  2.116 &  0.330 &  0.321 \\
0.010 & 7.3 & 0.0093 & 2.493 & 2.509 & 0.472 & 0.475 &  7.0 &  0.0362 & 
 2.061 &  2.069 &  0.277 & 0.267 \\
0.020 & 8.1 & 0.0161 & 2.425 & 2.429 & 0.377 & 0.371 &  7.9 &  0.0048 & 
 2.032 &  2.037 & 0.225 &  0.224 \\ \hline
 & \multicolumn{6}{c}{box CAP/EOM-EA-CCSD} & \multicolumn{6}{c}{ 
 projected box CAP/EOM-EA-CCSD, $n=9$ $^a$}\\
0.001 & 3.8 & 0.3603 & 2.212 & 2.301 & 0.422 & 0.271 &  3.8 &  0.3613 & 
 2.212 & 2.301 &  0.421 &  0.270 \\
0.005 & 6.2 & 0.0233 & 2.096 & 2.119 & 0.337 & 0.344 &  6.2 &  0.0244 & 
 2.095 &  2.119 &  0.335 & 0.342 \\
0.010 & 7.1 & 0.0156 & 2.059 & 2.066 & 0.286 & 0.290 &  7.1 &  0.0171 & 
 2.059 &  2.067 & 0.283 &  0.288 \\
0.020 & 7.9 & 0.0341 & 2.031 & 2.030 & 0.240 & 0.232 &  7.9 &  0.0326 & 
 2.031 &  2.031 & 0.238 &  0.230 \\ \hline
 & \multicolumn{6}{c}{smooth Voronoi CAP/EOM-EA-CCSD} \\
0.001 & 3.9 & 0.3228 & 2.209 & 2.308 & 0.400 & 0.272 \\
0.005 & 6.5 & 0.0203 & 2.081 & 2.095 & 0.308 & 0.314 \\
0.010 & 7.5 & 0.0025 & 2.043 & 2.045 & 0.250 & 0.250 \\
0.020 & 8.4 & 0.0224 & 2.014 & 2.012 & 0.199 & 0.195 \\ \hline
\multicolumn{3}{c}{Experiment\cite{walker78,panajotovic03}} & 1.8 & & 0.7 & \\ \hline
\end{tabular} \end{small} 

\footnotesize{$^a$ Using the $n$ lowest-lying right and left EOM-EA-CCSD eigenstates as basis.}
\label{tbl:c2h4}
\end{table}

As a second test case, we studied the $^2$B$_{2g}$ resonance of C$_2$H$_4^-$. Tab. 
\ref{tbl:c2h4} summarizes our results for the position and width of this resonance computed 
at the CAP-HF and CAP-EOM-EA-CCSD levels of theory. The corresponding $r^0$-trajectories 
are available in the Supporting Information. These trajectories look by and large similar to 
those of N$_2^-$ that we discussed in Secs. \ref{sec:n2hf} and \ref{sec:n2cc} with the following 
important differences: Firstly, we did not encounter any unphysical solutions of the CAP-HF 
equations at large $r^0$-values. Rather, the $r^0$-trajectories directly reach the CAP-free limit. 
Secondly, $\xi$ has multiple minima also in the CAP-HF case and, lastly, all CAP-HF trajectories 
show a discontinuity associated with an abrupt change in the character of the wave function. 

From the large $\xi$-values in Tab.~\ref{tbl:c2h4}, it is clear that 0.001 a.u. is too weak a CAP 
strength to describe the resonance state. Similar to N$_2^-$ (Tab.~\ref{tbl:n2}), increasing the 
CAP strength from 0.005 a.u. to 0.02 a.u. yields larger optimal $r^0$-values. An interesting 
difference to N$_2^-$ is, however, that the optimal $r^0$-values do not differ much between 
calculations with box CAP and smooth Voronoi CAP. 

The resonance positions obtained with CAP-EOM-EA-CCSD and CAP strengths of 0.005, 
0.01, and 0.02 a.u. differ by no more that 0.06 eV and are also in good agreement with 
previous CAP-EOM-EA-CCSD calculations in which $\eta$ was optimized at fixed CAP 
onset.\cite{zuev14,gayvert22b} In Ref. \citenum{zuev14}, the resonance position of 
C$_2$H$_4^-$ was computed to be 2.09 eV and 1.99 eV using the aug-cc-pVTZ+3s3p3d 
and aug-cc-pVQZ+3s3p3d basis sets, respectively. Notably, changing the CAP onset by 
0.5 a.u. led to changes of at most 0.02 eV in the resonance position of C$_2$H$_4^-$, 
i.e., the results from Ref. \citenum{zuev14} are less sensitive than those obtained with 
the present approach. CAP-HF yields considerably higher resonance positions than 
CAP-EOM-EA-CCSD, which is similar to N$_2^-$ (Tab. \ref{tbl:n2}) and can again be 
attributed to the neglect of electron correlation. Somewhat surprisingly, however, $\eta$ 
has a bigger impact on the resonance positions in the CAP-HF calculations. 

The CAP-EOM-EA-CCSD resonance widths computed with CAP strengths of 0.005 a.u., 
0.01 a.u., and 0.02 a.u. differ by up to 0.10 eV and are thus also more sensitive than in the 
approach where the CAP strength is optimized at fixed onset. In Ref. \citenum{zuev14}, a 
change of 0.5 a.u. in the CAP onset led to changes of at most 0.05 eV in the width of 
C$_2$H$_4^-$. Also, the widths obtained with the present approach are smaller by 
0.10--0.20 eV than those obtained with the aug-cc-pVTZ+3s3p3d basis and fixed onset 
in Ref. \citenum{zuev14}. Moreover, CAP-HF yields a larger resonance width than 
CAP-EOM-EA-CCSD according to Tab. \ref{tbl:c2h4}, which is in disagreement not 
only with the results for N$_2^-$ but also with previous results for other 
molecules.\cite{jagau17,white17}

Using $\xi$ as a quality measure, our best values for the resonance position and width are 
2.04 eV and 0.25 eV, respectively. With respect to the value extracted from C$_2$H$_4$-$e^-$ 
scattering experiments,\cite{walker78,panajotovic03} the resonance is thus 0.24 eV too high 
in energy. The width is significantly underestimated with respect to the experimental value of 
0.7 eV. In view of the results from Ref. \citenum{zuev14}, these discrepancies can be ascribed 
at least partly to our use of the relatively small cc-pVTZ+3p basis set. 

We note that the correction from Eq. \eqref{eq:depert2} is negligible as long as an adequate 
CAP strength is used, indicated by a small value of $\xi$. Differences between box and Voronoi 
CAPs are very small in CAP-EOM-EA-CCSD calculations but significant in CAP-HF calculations 
with the same $\eta$. If one compares results computed with those $\eta$-values that yield the 
smallest $\xi$ (0.02 a.u. for box CAP, 0.01 a.u. for Voronoi CAP), the agreement becomes much 
better for CAP-HF too. 

Regarding projected CAP-EOM-EA-CCSD, Tab. \ref{tbl:c2h4} shows that a basis of two CAP-free 
EOM-CCSD states is already sufficient to describe the position of C$_2$H$_4^-$ accurately, 
whereas the width is underestimated by up to 0.08 eV. With a basis of 5 states, the 
deviation in the width goes down to 0.02 eV and with a basis of 9 states, the widhts are essentially 
identical to those from full CAP-EOM-EA-CCSD.


\subsection{Anion of formic acid} \label{sec:hcooh}

\begin{table}[htb] \setlength{\tabcolsep}{6pt}
\caption{Vertical attachment energies $\Delta E$ and resonance widths $\Gamma$ in eV 
for the $^2$A'' resonance of HCOOH$^-$ computed with different CAP methods and the 
cc-pVTZ+3p basis set. $\Delta\widetilde{E}$ and $\widetilde{\Gamma}$ are the deperturbed 
vertical attachment energy and resonance width according to Eq. \eqref{eq:depert2}. Optimal 
values for $r^0$ according to Eq. \eqref{eq:xiopt} and corresponding values of $\xi$ are given 
in a.u.}
\begin{small} \begin{tabular}{lllllllllllll} \hline
$\eta$ & $r^0_\text{opt}$ & $\xi(r^0_\text{opt})$ & $\Delta E$ & $\Delta\widetilde{E}$ & 
$\Gamma$ & $\widetilde{\Gamma}$ & $r^0_\text{opt}$ & $\xi(r^0_\text{opt})$ & $\Delta E$ & 
$\Delta\widetilde{E}$ & $\Gamma$ & $\widetilde{\Gamma}$\\ \hline
 & \multicolumn{6}{c}{box CAP/HF/\textit{cis}-conformer} & \multicolumn{6}{c}{projected 
box CAP/EOM-EA-CCSD, $n=2$ $^a$} \\
0.001 & 4.1 & 0.0059 & 2.994 & 2.976 & 0.490 & 0.490 & \multicolumn{6}{l}{\textit{cis}-conformer} \\
0.005 & 6.6 & 0.0042 & 3.002 & 2.989 & 0.513 & 0.513 & 7.5 & 0.7206 & 1.825 & 1.921 & 0.597 & 0.168 \\
0.010 & 7.5 & 0.0047 & 3.006 & 2.994 & 0.525 & 0.524 & 8.8 & 0.5517 & 1.813 & 1.913 & 0.441 & 0.199 \\
0.020 & 8.2 & 0.0034 & 3.013 & 3.002 & 0.536 & 0.536 & 9.8 & 0.3517 & 1.793 & 1.893 & 0.372 & 0.243 \\
all & big & 0.0000 & 2.952 & 2.952 & 0.477 & 0.477 \\
all & big & 0.0000 & 2.710 & 2.710 & 0.044 & 0.044 \\ \hline
 & \multicolumn{6}{c}{smooth Voronoi CAP/HF/\textit{cis}-conformer} & 
\multicolumn{6}{c}{projected box CAP/EOM-EA-CCSD, $n=6$ $^a$} \\
0.001 & 4.9 & 0.0053 & 2.990 & 2.974 & 0.486 & 0.486 & \multicolumn{6}{l}{\textit{cis}-conformer} \\
0.005 & 7.1 & 0.0040 & 3.011 & 3.000 & 0.517 & 0.516 & 3.5 & 0.0365 & 2.656 & 2.753 & 0.700 & 0.702 \\
0.010 & 7.9 & 0.0038 & 3.020 & 3.011 & 0.532 & 0.533 & 4.1 & 0.0243 & 2.617 & 2.675 & 0.584 & 0.578 \\
0.020 & 8.8 & 0.0030 & 3.026 & 3.017 & 0.543 & 0.543 & 4.8 & 0.0287 & 2.588 & 2.628 & 0.501 & 0.513 \\
all & big & 0.0000 & 2.952 & 2.952 & 0.477 & 0.477 \\
all & big & 0.0000 & 2.710 & 2.710 & 0.044 & 0.044 \\ \hline 
 & \multicolumn{6}{c}{box CAP/EOM-EA-CCSD/\textit{cis}-conformer} & 
 \multicolumn{6}{c}{ box CAP/EOM-EA-CCSD/\textit{trans}-conformer} \\  
0.001 & 2.7 & 0.5095 & 1.794 & 1.957 & 2.021 & 1.008 & 1.9 & 0.3244 & 
 2.305 &  1.968 &  1.253 & 0.890 \\
0.005 & 3.2 & 0.0287 & 2.703 & 2.699 & 0.921 & 0.895 & 3.7 & 0.0428 & 
 2.367 &  2.286 &  0.515 & 0.528 \\
0.010 & 3.7 & 0.0265 & 2.696 & 2.651 & 0.754 & 0.770 & 4.2 & 0.0384 & 
 2.383 & 2.293 &  0.439 & 0.436 \\
0.020 & 4.2 & 0.0447 & 2.694 & 2.639 & 0.614 & 0.639 & 4.6 & 0.0393 & 
 2.408 &  2.319 &  0.390 & 0.395 \\ \hline
\multicolumn{8}{c}{smooth Voronoi CAP/EOM-EA-CCSD/\textit{cis}-conformer} \\ 
0.001 & 2.7 & 0.5025 & 1.813 & 1.927 & 1.812 & 0.909 \\				
0.005 & 2.5 & 0.0356 & 2.723 & 2.661 & 0.755 & 0.776 \\			
0.010 & 3.0 & 0.0322 & 2.737 & 2.653 & 0.629 & 0.635 \\
0.020 & 3.5 & 0.0398 & 2.747 & 2.659 & 0.523 & 0.510 \\ \hline
\multicolumn{3}{c}{Experiment\cite{aflatooni01}} & 1.73 &  &   &  \\ \hline
\end{tabular} \end{small}

\footnotesize{$^a$ Using the $n$ lowest-lying right and left EOM-EA-CCSD eigenstates as basis.}
\label{tbl:hcooh}
\end{table}

As a third test case, we studied the $^2$A'' resonance of the \textit{cis}-conformer of HCOOH$^-$. 
Positions and widths computed for this resonances with CAP-HF and CAP-EOM-EA-CCSD are 
summarized in Tab. \ref{tbl:hcooh}. The corresponding $r^0$-trajectories are available in the 
Supporting Information. The basic structure of these trajectories is the same as for N$_2^-$ 
and C$_2$H$_4^-$. At large $r^0$-values above 10 a.u., CAP-HF yields unphysical solutions 
similar to those discussed for N$_2^-$ in Sec. \ref{sec:n2hf}. Also, discontinuities are present 
in some CAP-EOM-EA-CCSD trajectories and $\xi$ has multiple minima in these cases. The 
resonance positions and widths reported for HCOOH$^-$ in Tab. \ref{tbl:hcooh} are evaluated 
at the global minimum of $\xi$.

Similar to the previously discussed molecules, 0.001 a.u. is again too weak a CAP strength in 
CAP-EOM-EA-CCSD calculations. Higher CAP strengths between 0.005 a.u. and 0.02 a.u. 
produce almost identical resonance positions with the variation not exceeding 0.02 eV, while 
the switch from box CAP to Voronoi CAP changes the position by no more than 0.04 eV. The 
impact of the correction from Eq. \eqref{eq:depert2} is in some cases a little larger than for 
N$_2^-$ and C$_2$H$_4^-$ but does not exceed 0.09 eV. CAP-HF yields significantly higher 
resonance positions than CAP-EOM-EA-CCSD owing to the neglect of electron correlation. 

Using $\xi$ as a quality measure, our best value for the resonance position of HCOOH$^-$ 
is 2.65 eV. An assessment of this value is complicated by the fact that previous theoretical 
investigations focused on the \textit{trans}-conformer of HCOOH, which is more stable than 
the \textit{cis}-conformer by 0.17 eV.\cite{hocking76} We therefore conducted 
additional CAP-EOM-EA-CCSD calculations on the \textit{trans}-conformer, which are reported 
in the lower right section of Table \ref{tbl:hcooh}. Our best value for the resonance position of 
the \textit{trans}-conformer is 2.29 eV, i.e., 0.36 eV less than for the \textit{cis}-conformer. This 
confirms previous results: With CAP-EOM-EA-CCSD, fixed CAP onset and optimized $\eta$, 
a value of 2.287 eV was computed for the resonance position at the \textit{trans}-structure,\cite{
benda18} while scattering calculations with the complex Kohn method yielded a position 
of ~1.9 eV.\cite{rescigno06} Electron transmission spectroscopy suggested a value of 1.73~eV 
for the position of HCOOH$^-$,\cite{aflatooni01} while the dissociative electron attachment 
cross section has a peak at 1.25~eV.\cite{pelc02} We also mention electron energy loss 
spectroscopy on HCOOH\cite{allan06} and measurements of the differential cross section 
in HCOOH-$e^-$ scattering.\cite{vizcaino06} Moreover, there is evidence that the anion 
potential energy surface is flatter than that of neutral HCOOH.\cite{rescigno06,benda18} 

Regarding the resonance width, Tab. \ref{tbl:hcooh} demonstrates that technical details make 
a sizable impact, especially in the case of CAP-EOM-EA-CCSD. Widths of the 
\textit{cis}-conformer obtained with CAP strengths of 0.005 a.u. and 0.02 a.u. differ by up to 
0.30 eV even though $\xi$ is of the same order of magnitude. Also, the width changes by up 
to 0.17 eV when going from box CAP to Voronoi CAP. The correction from Eq. \eqref{eq:depert2}, 
however, has a surprisingly small impact on the width not exceeding 0.03 eV. With CAP-HF 
the resonance is significantly narrower than with CAP-EOM-EA-CCSD similar to what we 
found for N$_2^-$. 

Our best values for the resonance widths of the \textit{cis}- and \textit{trans}-conformers 
are 0.770 eV and 0.436 eV, respectively. The latter value is considerably larger than 
the resonance width of 0.27 eV computed for the \textit{trans}-conformer\cite{benda18} 
with CAP-EOM-EA-CCSD, fixed CAP onset and optimized $\eta$. We add that the complex Kohn 
method yielded a value of 0.2 eV for the resonance width at the \textit{trans}-structure.\cite{
rescigno06} Notwithstanding the large variations among the $\Gamma$-values in Tab.~\ref{tbl:hcooh}, 
our results suggest that the \textit{cis}-conformer has a larger resonance width than the 
\textit{trans}-conformer.

As concerns projected CAP-EOM-EA-CCSD, Tab. \ref{tbl:hcooh} demonstrates that a basis 
of two states is not adequate for either the position or the width of HCOOH$^-$. Even with six 
states as basis, the discrepancy between full and projected CAP-EOM-EA-CCSD still amounts 
to 0.05--0.10 eV for the position and 0.11--0.22 eV for the width. 


\subsection{Role of diffuse basis functions} \label{sec:basis}

\begin{table}[htb] \setlength{\tabcolsep}{8pt}
\caption{Vertical attachment energies $\Delta E$ and resonance widths $\Gamma$ in eV for the 
$^2\Pi_g$ resonance of N$_2^-$ and the $^2$B$_{2g}$ resonance of C$_2$H$_4^-$ computed 
using different basis sets cc-pVTZ+$n$p and a CAP strength of 0.005 a.u. $\Delta\widetilde{E}$ 
and $\widetilde{\Gamma}$ are the deperturbed vertical attachment energy and resonance width 
according to Eq. \eqref{eq:depert2}. Optimal values for $r^0$ according to Eq. \eqref{eq:xiopt} 
and corresponding values of $\xi$ are given in a.u.}
\begin{tabular}{lllllll} \hline
$n$ & $r^0_\text{opt}$ & $\xi(r^0_\text{opt})$ & $\Delta E$ & $\Delta\widetilde{E}$ & $\Gamma$ 
& $\widetilde{\Gamma}$ \\ \hline
\multicolumn{7}{c}{N$_2^-$ / box CAP / HF} \\
3 & 4.6 & 0.0100 & 2.878 & 2.876 & 0.143 & 0.142 \\
6 & 7.8 & 0.0046 & 2.849 & 2.847 & 0.285 & 0.286 \\
8 & 7.8 & 0.0003 & 2.848 & 2.847 & 0.282 & 0.282 \\
12 & 7.8 & 0.0010 & 2.848 & 2.847 & 0.281 & 0.281 \\ \hline 
\multicolumn{3}{c}{Reference value\cite{berman83}} & 2.32 &  & 0.41 &  \\ \hline
\multicolumn{7}{c}{C$_2$H$_4^-$ / smooth Voronoi CAP / HF$^a$} \\
2 & 2.7 & 0.0290 & 2.987 & 2.901 & 0.742 & 0.743 \\
3 & 5.6 & 0.0105 & 2.791 & 2.818 & 0.381 & 0.380 \\
4 & 4.1 & 0.0076 & 2.891 & 2.872 & 0.433 & 0.435 \\ \hline 
\multicolumn{7}{c}{C$_2$H$_4^-$ / smooth Voronoi CAP/EOM-EA-CCSD$^a$} \\
2 & 2.2 & 0.0927 & 2.587 & 2.368 & 1.048 & 1.008 \\
3 & 5.3 & 0.0022 & 2.218 & 2.222 & 0.292 & 0.291 \\
4 & 4.3 & 0.0230 & 2.311 & 2.270 & 0.402 & 0.396 \\
5 & 4.1 & 0.0216 & 2.332 & 2.283 & 0.422 & 0.423 \\
6 & 3.9 & 0.0286 & 2.353 & 2.293 & 0.446 & 0.440 \\
7 & 3.8 & 0.0300 & 2.364 & 2.297 & 0.464 & 0.459 \\
8 & 3.8 & 0.0288 & 2.364 & 2.297 & 0.464 & 0.461 \\ \hline 
\multicolumn{3}{c}{Experiment\cite{walker78,panajotovic03}} & 1.8 &  & 0.7 &  \\ \hline
\end{tabular} 
\label{tbl:basis}
\end{table}

It is well established that the description of TAs by means of CAP methods mandates the use 
of customized basis sets.\cite{jagau17} A possible strategy is to augment correlation consistent 
basis sets\cite{dunning89,kendall92} by additional shells of diffuse functions. Whereas one can 
anticipate that $p$-shells are most important to describe $\pi^*$-resonances such as the ones 
studied here, the number of shells required for convergence varies from case to case. 

In Tab. \ref{tbl:basis}, we report resonance positions and widths for N$_2^-$ and C$_2$H$_4^-$ 
that were computed with different numbers of diffuse shells included in the basis set. For N$_2^-$, 
the additional shells are even tempered, whereas this was not possible for C$_2$H$_4^-$ because 
of linear dependencies in the basis. For the latter molecule, the exponents of the additional shells 
are available in the Supporting Information. 

The upper section of Tab. \ref{tbl:basis} shows that basis set convergence is achieved relatively 
easily in CAP-HF calculations on N$_2^-$. The resonance position is accurate to 0.03 eV already 
with three extra $p$-shells; and with six of them the resonance width is converged as well. The 
fact that more than six extra $p$-shells are not needed is also apparent from the $r^0$-trajectories 
reported in the Supporting Information, which barely change upon enlarging the basis further. 

It is noteworthy that even in a basis set with twelve additional $p$-shells where the smallest 
exponent is of the order of $10^{-5}$, convergence of the CAP-HF equations to the desired 
solution can still be achieved for N$_2^-$. This is not the case for C$_2$H$_4^-$ where we 
were unable to find the CAP-HF solution corresponding to the $^2$B$_{2g}$ resonance when 
more than four diffuse $p$-shells were present. The middle section of Tab. \ref{tbl:basis} 
illustrates that the position and width of C$_2$H$_4^-$ are not yet converged in this basis. 

With CAP-EOM-EA-CCSD, such convergence problems does not arise because the CAP-HF 
equations are solved for the neutral molecule. Although it is not always trivial to identify the 
resonance state among the CAP-EOM-EA-CCSD solutions, we were able to identify the correct 
state of C$_2$H$_4^-$ in basis sets with up to eight extra $p$-shells. The lower part of Tab. 
\ref{tbl:basis} illustrates that with six $p$-shells the position and width are converged to within 
0.01 eV and 0.02 eV, respectively.


\section{Conclusions}\label{sec:conc}

We have introduced a method for optimizing the onset of complex absorbing potentials that  
offer the possibility to compute energies and widths of molecular temporary anions in a basis set 
of atom-centered Gaussian functions. Our method is based on an error function that evaluates 
what fraction of the resonance position and width are due to the expectation value of the CAP. 
If the values of these fractions are close to one, the CAP exclusively acts as a perturbation and 
does not stabilize the resonance. Conversely, the closer the values are to zero, the less perturbed 
the resonance wave function is by the CAP. The new criterion can be viewed as a generalization 
of the established criterion for optimizing the CAP strength and can, in principle, be further 
generalized to arbitrary CAP parameters. 

We tested our new approach for the $\pi^*$ temporary anions of N$_2$, C$_2$H$_4$, and HCOOH 
using box CAPs and smooth Voronoi CAPs at the HF and EOM-EA-CCSD levels of theory. We also 
used the projected CAP-EOM-EA-CCSD method where the CAP is represented in a basis of CAP-free 
EOM-EA-CCSD states. The optimal onsets usually vary only little between box and smooth Voronoi 
CAP as well as between HF, full and projected EOM-EA-CCSD calculations. Overall, the performance 
of the new approach is similar to that of the established one in which the CAP strength is optimized 
although we found substantial discrepancies in the resonance width in some cases. 

The computational cost of both approaches is ultimately comparable. Whereas the established 
technique entails the need to run series of calculations with different CAP strength, our new technique 
requires to run analogous series in which the onset parameters are varied. This equally applies to 
full and projected CAP calculations. Also, using a large enough CAP strength is important in the new 
approach just like choosing the onset properly is important in the established approach. Nonetheless, 
we believe that optimization of the CAP onset as proposed in the present work will prove useful to 
resolve pathological cases where optimization of only the CAP strength does not yield satisfactory 
results. 


\begin{acknowledgement}
Funding from the European Research Council (ERC) under the European Union's Horizon 2020 research 
and innovation program (Grant Agreement No. 851766) and the KU Leuven internal funds (Grant No. 
C14/22/083) is gratefully acknowledged. Resources and services used in this work were partly provided 
by the VSC (Flemish Supercomputer Center) funded by the Research Foundation---Flanders (FWO) and 
the Flemish Government.
\end{acknowledgement}

\bibliography{CAPbox}

\providecommand{\latin}[1]{#1}
\makeatletter
\providecommand{\doi}
  {\begingroup\let\do\@makeother\dospecials
  \catcode`\{=1 \catcode`\}=2 \doi@aux}
\providecommand{\doi@aux}[1]{\endgroup\texttt{#1}}
\makeatother
\providecommand*\mcitethebibliography{\thebibliography}
\csname @ifundefined\endcsname{endmcitethebibliography}
  {\let\endmcitethebibliography\endthebibliography}{}
\begin{mcitethebibliography}{71}
\providecommand*\natexlab[1]{#1}
\providecommand*\mciteSetBstSublistMode[1]{}
\providecommand*\mciteSetBstMaxWidthForm[2]{}
\providecommand*\mciteBstWouldAddEndPuncttrue
  {\def\EndOfBibitem{\unskip.}}
\providecommand*\mciteBstWouldAddEndPunctfalse
  {\let\EndOfBibitem\relax}
\providecommand*\mciteSetBstMidEndSepPunct[3]{}
\providecommand*\mciteSetBstSublistLabelBeginEnd[3]{}
\providecommand*\EndOfBibitem{}
\mciteSetBstSublistMode{f}
\mciteSetBstMaxWidthForm{subitem}{(\alph{mcitesubitemcount})}
\mciteSetBstSublistLabelBeginEnd
  {\mcitemaxwidthsubitemform\space}
  {\relax}
  {\relax}

\bibitem[Bouda\"{i}ffa \latin{et~al.}(2000)Bouda\"{i}ffa, Cloutier, Hunting,
  Huels, and Sanche]{boudaiffa00}
Bouda\"{i}ffa,~B.; Cloutier,~P.; Hunting,~D.; Huels,~M.~A.; Sanche,~L. Resonant
  Formation of {DNA} Strand Breaks by Low-Energy (3 to 20 eV) Electrons.
  \emph{Science} \textbf{2000}, \emph{287}, 1658--1660\relax
\mciteBstWouldAddEndPuncttrue
\mciteSetBstMidEndSepPunct{\mcitedefaultmidpunct}
{\mcitedefaultendpunct}{\mcitedefaultseppunct}\relax
\EndOfBibitem
\bibitem[Simons(2007)]{simons07}
Simons,~J. How Very Low-Energy (0.1--2 eV) Electrons Cause {DNA} Strand Breaks.
  \emph{Adv. Quantum Chem.} \textbf{2007}, \emph{52}, 171--188\relax
\mciteBstWouldAddEndPuncttrue
\mciteSetBstMidEndSepPunct{\mcitedefaultmidpunct}
{\mcitedefaultendpunct}{\mcitedefaultseppunct}\relax
\EndOfBibitem
\bibitem[Millar \latin{et~al.}(2017)Millar, Walsh, and Field]{millar17}
Millar,~T.~J.; Walsh,~C.; Field,~T.~A. Negative Ions in Space. \emph{Chem.
  Rev.} \textbf{2017}, \emph{117}, 1765--1795\relax
\mciteBstWouldAddEndPuncttrue
\mciteSetBstMidEndSepPunct{\mcitedefaultmidpunct}
{\mcitedefaultendpunct}{\mcitedefaultseppunct}\relax
\EndOfBibitem
\bibitem[Stoffels \latin{et~al.}(2001)Stoffels, Stoffels, and
  Kroesen]{stoffels01}
Stoffels,~E.; Stoffels,~W.~W.; Kroesen,~G. M.~W. Plasma chemistry and surface
  processes of negative ions. \emph{Plasma Sources Sci. Technol.}
  \textbf{2001}, \emph{10}, 311--317\relax
\mciteBstWouldAddEndPuncttrue
\mciteSetBstMidEndSepPunct{\mcitedefaultmidpunct}
{\mcitedefaultendpunct}{\mcitedefaultseppunct}\relax
\EndOfBibitem
\bibitem[Herbert(2015)]{herbert15}
Herbert,~J.~M. The quantum chemistry of loosely-bound electrons. \emph{Rev.
  Comp. Chem.} \textbf{2015}, \emph{28}, 391--517\relax
\mciteBstWouldAddEndPuncttrue
\mciteSetBstMidEndSepPunct{\mcitedefaultmidpunct}
{\mcitedefaultendpunct}{\mcitedefaultseppunct}\relax
\EndOfBibitem
\bibitem[Ing{\'o}lfsson(2019)]{ingolfsson19}
Ing{\'o}lfsson,~O., Ed. \emph{Low-Energy Electrons: Fundamentals and
  Applications}; Pan Stanford Publishing, 2019\relax
\mciteBstWouldAddEndPuncttrue
\mciteSetBstMidEndSepPunct{\mcitedefaultmidpunct}
{\mcitedefaultendpunct}{\mcitedefaultseppunct}\relax
\EndOfBibitem
\bibitem[Moiseyev(2011)]{moiseyev11}
Moiseyev,~N. \emph{Non-Hermitian Quantum Mechanics}; Cambridge University
  Press, 2011\relax
\mciteBstWouldAddEndPuncttrue
\mciteSetBstMidEndSepPunct{\mcitedefaultmidpunct}
{\mcitedefaultendpunct}{\mcitedefaultseppunct}\relax
\EndOfBibitem
\bibitem[Jagau \latin{et~al.}(2017)Jagau, Bravaya, and Krylov]{jagau17}
Jagau,~T.-C.; Bravaya,~K.~B.; Krylov,~A.~I. Extending Quantum Chemistry of
  Bound States to Electronic Resonances. \emph{Annu. Rev. Phys. Chem.}
  \textbf{2017}, \emph{68}, 525--553\relax
\mciteBstWouldAddEndPuncttrue
\mciteSetBstMidEndSepPunct{\mcitedefaultmidpunct}
{\mcitedefaultendpunct}{\mcitedefaultseppunct}\relax
\EndOfBibitem
\bibitem[Jagau(2022)]{jagau22}
Jagau,~T.~C. Theory of electronic resonances: Fundamental aspects and recent
  advances. \emph{Chem. Commun.} \textbf{2022}, \emph{58}, 5205--5224\relax
\mciteBstWouldAddEndPuncttrue
\mciteSetBstMidEndSepPunct{\mcitedefaultmidpunct}
{\mcitedefaultendpunct}{\mcitedefaultseppunct}\relax
\EndOfBibitem
\bibitem[Bardsley and Mandl(1968)Bardsley, and Mandl]{bardsley68}
Bardsley,~J.~N.; Mandl,~F. Resonant scattering of electrons by molecules.
  \emph{Rep. Prog. Phys.} \textbf{1968}, \emph{31}, 471--531\relax
\mciteBstWouldAddEndPuncttrue
\mciteSetBstMidEndSepPunct{\mcitedefaultmidpunct}
{\mcitedefaultendpunct}{\mcitedefaultseppunct}\relax
\EndOfBibitem
\bibitem[Taylor(1971)]{taylor71}
Taylor,~J.~R. \emph{Scattering Theory: The Quantum Theory on Nonrelativistic
  Collisions}; Wiley, 1971\relax
\mciteBstWouldAddEndPuncttrue
\mciteSetBstMidEndSepPunct{\mcitedefaultmidpunct}
{\mcitedefaultendpunct}{\mcitedefaultseppunct}\relax
\EndOfBibitem
\bibitem[Aguilar and Combes(1971)Aguilar, and Combes]{aguilar71}
Aguilar,~J.; Combes,~J.~M. A class of analytic perturbations for one-body
  Schr\"odinger Hamiltonians. \emph{Commun. Math. Phys.} \textbf{1971},
  \emph{22}, 269--279\relax
\mciteBstWouldAddEndPuncttrue
\mciteSetBstMidEndSepPunct{\mcitedefaultmidpunct}
{\mcitedefaultendpunct}{\mcitedefaultseppunct}\relax
\EndOfBibitem
\bibitem[Balslev and Combes(1971)Balslev, and Combes]{balslev1971}
Balslev,~E.; Combes,~J.~M. Spectral properties of many-body Schr\"odinger
  operators with dilatation-analytic interactions. \emph{Commun. Math. Phys.}
  \textbf{1971}, \emph{22}, 280--294\relax
\mciteBstWouldAddEndPuncttrue
\mciteSetBstMidEndSepPunct{\mcitedefaultmidpunct}
{\mcitedefaultendpunct}{\mcitedefaultseppunct}\relax
\EndOfBibitem
\bibitem[Moiseyev(1998)]{moiseyev98a}
Moiseyev,~N. Quantum theory of resonances: calculating energies, widths and
  cross-sections by complex scaling. \emph{Phys. Rep.} \textbf{1998},
  \emph{302}, 212--293\relax
\mciteBstWouldAddEndPuncttrue
\mciteSetBstMidEndSepPunct{\mcitedefaultmidpunct}
{\mcitedefaultendpunct}{\mcitedefaultseppunct}\relax
\EndOfBibitem
\bibitem[McCurdy and Rescigno(1978)McCurdy, and Rescigno]{mccurdy78}
McCurdy,~C.~W.; Rescigno,~T.~N. Extension of the Method of Complex Basis
  Functions to Molecular Resonances. \emph{Phys. Rev. Lett.} \textbf{1978},
  \emph{41}, 1364--1368\relax
\mciteBstWouldAddEndPuncttrue
\mciteSetBstMidEndSepPunct{\mcitedefaultmidpunct}
{\mcitedefaultendpunct}{\mcitedefaultseppunct}\relax
\EndOfBibitem
\bibitem[Moiseyev and Corcoran(1979)Moiseyev, and Corcoran]{moiseyev79}
Moiseyev,~N.; Corcoran,~C. Autoionizing states of {H}$_2$ and {H}$_2^-$ using
  the complex-scaling method. \emph{Phys. Rev. A} \textbf{1979}, \emph{20},
  814--817\relax
\mciteBstWouldAddEndPuncttrue
\mciteSetBstMidEndSepPunct{\mcitedefaultmidpunct}
{\mcitedefaultendpunct}{\mcitedefaultseppunct}\relax
\EndOfBibitem
\bibitem[White \latin{et~al.}(2015)White, Head-Gordon, and McCurdy]{white15}
White,~A.~F.; Head-Gordon,~M.; McCurdy,~C.~W. Complex basis functions
  revisited: Implementation with applications to carbon tetrafluoride and
  aromatic N-containing heterocycles within the static-exchange approximation.
  \emph{J. Chem. Phys.} \textbf{2015}, \emph{142}, 054103\relax
\mciteBstWouldAddEndPuncttrue
\mciteSetBstMidEndSepPunct{\mcitedefaultmidpunct}
{\mcitedefaultendpunct}{\mcitedefaultseppunct}\relax
\EndOfBibitem
\bibitem[Jolicard and Austin(1985)Jolicard, and Austin]{jolicard85}
Jolicard,~G.; Austin,~E.~J. Optical potential stabilization method for
  predicting resonance levels. \emph{Chem. Phys. Lett.} \textbf{1985},
  \emph{121}, 106--110\relax
\mciteBstWouldAddEndPuncttrue
\mciteSetBstMidEndSepPunct{\mcitedefaultmidpunct}
{\mcitedefaultendpunct}{\mcitedefaultseppunct}\relax
\EndOfBibitem
\bibitem[Jolicard and Austin(1986)Jolicard, and Austin]{jolicard86}
Jolicard,~G.; Austin,~E.~J. Optical potential method of calculating resonance
  energies and widths. \emph{Chem. Phys.} \textbf{1986}, \emph{103},
  295--302\relax
\mciteBstWouldAddEndPuncttrue
\mciteSetBstMidEndSepPunct{\mcitedefaultmidpunct}
{\mcitedefaultendpunct}{\mcitedefaultseppunct}\relax
\EndOfBibitem
\bibitem[Riss and Meyer(1993)Riss, and Meyer]{riss93}
Riss,~U.~V.; Meyer,~H.-D. Calculation of resonance energies and widths using
  the complex absorbing potential method. \emph{J. Phys. B: At. Mol. Opt.
  Phys.} \textbf{1993}, \emph{26}, 4503--4535\relax
\mciteBstWouldAddEndPuncttrue
\mciteSetBstMidEndSepPunct{\mcitedefaultmidpunct}
{\mcitedefaultendpunct}{\mcitedefaultseppunct}\relax
\EndOfBibitem
\bibitem[Muga \latin{et~al.}(2004)Muga, Palao, Navarro, and Egusquiza]{muga04}
Muga,~J.~G.; Palao,~J.~P.; Navarro,~B.; Egusquiza,~I.~L. Complex absorbing
  potentials. \emph{Phys. Rep.} \textbf{2004}, \emph{395}, 357--426\relax
\mciteBstWouldAddEndPuncttrue
\mciteSetBstMidEndSepPunct{\mcitedefaultmidpunct}
{\mcitedefaultendpunct}{\mcitedefaultseppunct}\relax
\EndOfBibitem
\bibitem[Moiseyev \latin{et~al.}(1978)Moiseyev, Certain, and
  Weinhold]{moiseyev78}
Moiseyev,~N.; Certain,~P.~R.; Weinhold,~F. Resonance properties of
  complex-rotated {H}amiltonians. \emph{Mol. Phys.} \textbf{1978}, \emph{36},
  1613--1630\relax
\mciteBstWouldAddEndPuncttrue
\mciteSetBstMidEndSepPunct{\mcitedefaultmidpunct}
{\mcitedefaultendpunct}{\mcitedefaultseppunct}\relax
\EndOfBibitem
\bibitem[Siegert(1939)]{siegert39}
Siegert,~A. J.~F. On the Derivation of the Dispersion Formula for Nuclear
  Reactions. \emph{Phys. Rev.} \textbf{1939}, \emph{56}, 750--752\relax
\mciteBstWouldAddEndPuncttrue
\mciteSetBstMidEndSepPunct{\mcitedefaultmidpunct}
{\mcitedefaultendpunct}{\mcitedefaultseppunct}\relax
\EndOfBibitem
\bibitem[Kosloff and Kosloff(1986)Kosloff, and Kosloff]{kosloff86}
Kosloff,~R.; Kosloff,~D. Absorbing boundaries for wave propagation problems.
  \emph{J. Comput. Phys.} \textbf{1986}, \emph{63}, 363--376\relax
\mciteBstWouldAddEndPuncttrue
\mciteSetBstMidEndSepPunct{\mcitedefaultmidpunct}
{\mcitedefaultendpunct}{\mcitedefaultseppunct}\relax
\EndOfBibitem
\bibitem[Neuhauser and Baer(1989)Neuhauser, and Baer]{neuhauser89a}
Neuhauser,~D.; Baer,~M. The time-dependent {S}chr\"odinger equation:
  {A}pplication of absorbing boundary conditions. \emph{J. Chem. Phys.}
  \textbf{1989}, \emph{90}, 4351--4355\relax
\mciteBstWouldAddEndPuncttrue
\mciteSetBstMidEndSepPunct{\mcitedefaultmidpunct}
{\mcitedefaultendpunct}{\mcitedefaultseppunct}\relax
\EndOfBibitem
\bibitem[Neuhauser and Baer(1989)Neuhauser, and Baer]{neuhauser89b}
Neuhauser,~D.; Baer,~M. The application of wave packets to reactive atom-diatom
  systems: A new approach. \emph{J. Chem. Phys.} \textbf{1989}, \emph{91},
  4651--4657\relax
\mciteBstWouldAddEndPuncttrue
\mciteSetBstMidEndSepPunct{\mcitedefaultmidpunct}
{\mcitedefaultendpunct}{\mcitedefaultseppunct}\relax
\EndOfBibitem
\bibitem[Mac\'ias \latin{et~al.}(1994)Mac\'ias, Brouard, and Muga]{macias94}
Mac\'ias,~D.; Brouard,~S.; Muga,~J. Optimization of absorbing potentials.
  \emph{Chem. Phys. Lett.} \textbf{1994}, \emph{228}, 672--677\relax
\mciteBstWouldAddEndPuncttrue
\mciteSetBstMidEndSepPunct{\mcitedefaultmidpunct}
{\mcitedefaultendpunct}{\mcitedefaultseppunct}\relax
\EndOfBibitem
\bibitem[Riss and Meyer(1995)Riss, and Meyer]{riss95}
Riss,~U.~V.; Meyer,~H.-D. Reflection-free complex absorbing potentials.
  \emph{J. Phys. B: At. Mol. Opt. Phys.} \textbf{1995}, \emph{28},
  1475--1493\relax
\mciteBstWouldAddEndPuncttrue
\mciteSetBstMidEndSepPunct{\mcitedefaultmidpunct}
{\mcitedefaultendpunct}{\mcitedefaultseppunct}\relax
\EndOfBibitem
\bibitem[Riss and Meyer(1996)Riss, and Meyer]{riss96}
Riss,~U.~V.; Meyer,~H.-D. Investigation on the reflection and transmission
  properties of complex absorbing potentials. \emph{J. Chem. Phys.}
  \textbf{1996}, \emph{105}, 1409--1419\relax
\mciteBstWouldAddEndPuncttrue
\mciteSetBstMidEndSepPunct{\mcitedefaultmidpunct}
{\mcitedefaultendpunct}{\mcitedefaultseppunct}\relax
\EndOfBibitem
\bibitem[Riss and Meyer(1998)Riss, and Meyer]{riss98}
Riss,~U.~V.; Meyer,~H.-D. The transformative complex absorbing potential
  method: a bridge between complex absorbing potentials and smooth exterior
  scaling. \emph{J. Phys. B: At. Mol. Opt. Phys.} \textbf{1998}, \emph{31},
  2279--2304\relax
\mciteBstWouldAddEndPuncttrue
\mciteSetBstMidEndSepPunct{\mcitedefaultmidpunct}
{\mcitedefaultendpunct}{\mcitedefaultseppunct}\relax
\EndOfBibitem
\bibitem[Moiseyev(1998)]{moiseyev98b}
Moiseyev,~N. Derivations of universal exact complex absorption potentials by
  the generalized complex coordinate method. \emph{J. Phys. B: At. Mol. Opt.
  Phys.} \textbf{1998}, \emph{31}, 1431--1441\relax
\mciteBstWouldAddEndPuncttrue
\mciteSetBstMidEndSepPunct{\mcitedefaultmidpunct}
{\mcitedefaultendpunct}{\mcitedefaultseppunct}\relax
\EndOfBibitem
\bibitem[Sommerfeld and Ehara(2015)Sommerfeld, and Ehara]{sommerfeld15}
Sommerfeld,~T.; Ehara,~M. Complex Absorbing Potentials with {Voronoi}
  Isosurfaces Wrapping Perfectly around Molecules. \emph{J. Chem. Theory
  Comput.} \textbf{2015}, \emph{11}, 4627--4633\relax
\mciteBstWouldAddEndPuncttrue
\mciteSetBstMidEndSepPunct{\mcitedefaultmidpunct}
{\mcitedefaultendpunct}{\mcitedefaultseppunct}\relax
\EndOfBibitem
\bibitem[Gyamfi and Jagau(2022)Gyamfi, and Jagau]{gyamfi22}
Gyamfi,~J.~A.; Jagau,~T.-C. Ab Initio Molecular Dynamics of Temporary Anions
  Using Complex Absorbing Potentials. \emph{J. Phys. Chem. Lett.}
  \textbf{2022}, \emph{13}, 8477--8483\relax
\mciteBstWouldAddEndPuncttrue
\mciteSetBstMidEndSepPunct{\mcitedefaultmidpunct}
{\mcitedefaultendpunct}{\mcitedefaultseppunct}\relax
\EndOfBibitem
\bibitem[Zhou and Ernzerhof(2012)Zhou, and Ernzerhof]{zhou12}
Zhou,~Y.; Ernzerhof,~M. Calculating the Lifetimes of Metastable States with
  Complex Density Functional Theory. \emph{J. Phys. Chem. Lett.} \textbf{2012},
  \emph{3}, 1916--1920\relax
\mciteBstWouldAddEndPuncttrue
\mciteSetBstMidEndSepPunct{\mcitedefaultmidpunct}
{\mcitedefaultendpunct}{\mcitedefaultseppunct}\relax
\EndOfBibitem
\bibitem[Jagau \latin{et~al.}(2014)Jagau, Zuev, Bravaya, Epifanovsky, and
  Krylov]{jagau14}
Jagau,~T.-C.; Zuev,~D.; Bravaya,~K.~B.; Epifanovsky,~E.; Krylov,~A.~I. A Fresh
  Look at Resonances and Complex Absorbing Potentials: Density Matrix-Based
  Approach. \emph{J. Phys. Chem. Lett.} \textbf{2014}, \emph{5}, 310--315\relax
\mciteBstWouldAddEndPuncttrue
\mciteSetBstMidEndSepPunct{\mcitedefaultmidpunct}
{\mcitedefaultendpunct}{\mcitedefaultseppunct}\relax
\EndOfBibitem
\bibitem[Zuev \latin{et~al.}(2014)Zuev, Jagau, Bravaya, Epifanovsky, Shao,
  Sundstrom, Head-Gordon, and Krylov]{zuev14}
Zuev,~D.; Jagau,~T.-C.; Bravaya,~K.~B.; Epifanovsky,~E.; Shao,~Y.;
  Sundstrom,~E.; Head-Gordon,~M.; Krylov,~A.~I. Complex absorbing potentials
  within {EOM-CC} family of methods: Theory, implementation, and benchmarks.
  \emph{J. Chem. Phys.} \textbf{2014}, \emph{141}, 024102\relax
\mciteBstWouldAddEndPuncttrue
\mciteSetBstMidEndSepPunct{\mcitedefaultmidpunct}
{\mcitedefaultendpunct}{\mcitedefaultseppunct}\relax
\EndOfBibitem
\bibitem[Benda and Jagau(2017)Benda, and Jagau]{benda17}
Benda,~Z.; Jagau,~T.-C. Analytic gradients for the complex absorbing potential
  equation-of-motion coupled-cluster method. \emph{J. Chem. Phys.}
  \textbf{2017}, \emph{146}, 031101\relax
\mciteBstWouldAddEndPuncttrue
\mciteSetBstMidEndSepPunct{\mcitedefaultmidpunct}
{\mcitedefaultendpunct}{\mcitedefaultseppunct}\relax
\EndOfBibitem
\bibitem[Sommerfeld \latin{et~al.}(1998)Sommerfeld, Riss, Meyer, Cederbaum,
  Engels, and Suter]{sommerfeld98}
Sommerfeld,~T.; Riss,~U.~V.; Meyer,~H.-D.; Cederbaum,~L.~S.; Engels,~B.;
  Suter,~H.~U. Temporary anions—calculation of energy and lifetime by
  absorbing potentials: the {N$_2^-$} {$^2\Pi_g$} resonance. \emph{J. Phys. B:
  At. Mol. Opt. Phys.} \textbf{1998}, \emph{31}, 4107--4122\relax
\mciteBstWouldAddEndPuncttrue
\mciteSetBstMidEndSepPunct{\mcitedefaultmidpunct}
{\mcitedefaultendpunct}{\mcitedefaultseppunct}\relax
\EndOfBibitem
\bibitem[Sommerfeld and Santra(2001)Sommerfeld, and Santra]{sommerfeld01}
Sommerfeld,~T.; Santra,~R. Efficient method to perform {CAP/CI} calculations
  for temporary anions. \emph{Int. J. Quantum Chem.} \textbf{2001}, \emph{82},
  218--226\relax
\mciteBstWouldAddEndPuncttrue
\mciteSetBstMidEndSepPunct{\mcitedefaultmidpunct}
{\mcitedefaultendpunct}{\mcitedefaultseppunct}\relax
\EndOfBibitem
\bibitem[Feuerbacher \latin{et~al.}(2003)Feuerbacher, Sommerfeld, Santra, and
  Cederbaum]{feuerbacher03}
Feuerbacher,~S.; Sommerfeld,~T.; Santra,~R.; Cederbaum,~L.~S. Complex absorbing
  potentials in the framework of electron propagator theory. {II.} Application
  to temporary anions. \emph{J. Chem. Phys.} \textbf{2003}, \emph{118},
  6188--6199\relax
\mciteBstWouldAddEndPuncttrue
\mciteSetBstMidEndSepPunct{\mcitedefaultmidpunct}
{\mcitedefaultendpunct}{\mcitedefaultseppunct}\relax
\EndOfBibitem
\bibitem[Sajeev \latin{et~al.}(2005)Sajeev, Santra, and Pal]{sajeev05}
Sajeev,~Y.; Santra,~R.; Pal,~S. Analytically continued {Fock} space
  multireference coupled-cluster theory: Application to the {$^2\Pi_g$} shape
  resonance in {e-N$_2$} scattering. \emph{J. Chem. Phys.} \textbf{2005},
  \emph{122}, 234320\relax
\mciteBstWouldAddEndPuncttrue
\mciteSetBstMidEndSepPunct{\mcitedefaultmidpunct}
{\mcitedefaultendpunct}{\mcitedefaultseppunct}\relax
\EndOfBibitem
\bibitem[Pal \latin{et~al.}(2006)Pal, Sajeev, and Vaval]{pal06}
Pal,~S.; Sajeev,~Y.; Vaval,~N. Analytically continued {Fock} space
  multi-reference coupled-cluster theory: {Application} to the shape resonance.
  \emph{Chem. Phys.} \textbf{2006}, \emph{329}, 283--289\relax
\mciteBstWouldAddEndPuncttrue
\mciteSetBstMidEndSepPunct{\mcitedefaultmidpunct}
{\mcitedefaultendpunct}{\mcitedefaultseppunct}\relax
\EndOfBibitem
\bibitem[Ghosh \latin{et~al.}(2012)Ghosh, Vaval, and Pal]{ghosh12}
Ghosh,~A.; Vaval,~N.; Pal,~S. Equation-of-motion coupled-cluster method for the
  study of shape resonance. \emph{J. Chem. Phys.} \textbf{2012}, \emph{136},
  234110\relax
\mciteBstWouldAddEndPuncttrue
\mciteSetBstMidEndSepPunct{\mcitedefaultmidpunct}
{\mcitedefaultendpunct}{\mcitedefaultseppunct}\relax
\EndOfBibitem
\bibitem[Jagau and Krylov(2014)Jagau, and Krylov]{jagau14b}
Jagau,~T.-C.; Krylov,~A.~I. Complex Absorbing Potential Equation-of-Motion
  Coupled-Cluster Method Yields Smooth and Internally Consistent Potential
  Energy Surfaces and Lifetimes for Molecular Resonances. \emph{J. Phys. Chem.
  Lett.} \textbf{2014}, \emph{5}, 3078--3085\relax
\mciteBstWouldAddEndPuncttrue
\mciteSetBstMidEndSepPunct{\mcitedefaultmidpunct}
{\mcitedefaultendpunct}{\mcitedefaultseppunct}\relax
\EndOfBibitem
\bibitem[Kunitsa \latin{et~al.}(2017)Kunitsa, Granovsky, and
  Bravaya]{kunitsa17}
Kunitsa,~A.~A.; Granovsky,~A.~A.; Bravaya,~K.~B. {CAP-XMCQDPT2} method for
  molecular electronic resonances. \emph{J. Chem. Phys.} \textbf{2017},
  \emph{146}, 184107\relax
\mciteBstWouldAddEndPuncttrue
\mciteSetBstMidEndSepPunct{\mcitedefaultmidpunct}
{\mcitedefaultendpunct}{\mcitedefaultseppunct}\relax
\EndOfBibitem
\bibitem[Benda \latin{et~al.}(2018)Benda, Rickmeyer, and Jagau]{benda18}
Benda,~Z.; Rickmeyer,~K.; Jagau,~T.-C. Structure Optimization of Temporary
  Anions. \emph{J. Chem. Theory Comput.} \textbf{2018}, \emph{14},
  3468--3478\relax
\mciteBstWouldAddEndPuncttrue
\mciteSetBstMidEndSepPunct{\mcitedefaultmidpunct}
{\mcitedefaultendpunct}{\mcitedefaultseppunct}\relax
\EndOfBibitem
\bibitem[Benda and Jagau(2018)Benda, and Jagau]{benda18b}
Benda,~Z.; Jagau,~T.-C. Understanding Processes Following Resonant Electron
  Attachment: Minimum-Energy Crossing Points between Anionic and Neutral
  Potential Energy Surfaces. \emph{J. Chem. Theory Comput.} \textbf{2018},
  \emph{14}, 4216--4223\relax
\mciteBstWouldAddEndPuncttrue
\mciteSetBstMidEndSepPunct{\mcitedefaultmidpunct}
{\mcitedefaultendpunct}{\mcitedefaultseppunct}\relax
\EndOfBibitem
\bibitem[Thodika \latin{et~al.}(2019)Thodika, Fennimore, Karsili, and
  Matsika]{thodika19}
Thodika,~M.; Fennimore,~M.; Karsili,~T. N.~V.; Matsika,~S. Comparative study of
  methodologies for calculating metastable states of small to medium-sized
  molecules. \emph{J. Chem. Phys.} \textbf{2019}, \emph{151}, 244104\relax
\mciteBstWouldAddEndPuncttrue
\mciteSetBstMidEndSepPunct{\mcitedefaultmidpunct}
{\mcitedefaultendpunct}{\mcitedefaultseppunct}\relax
\EndOfBibitem
\bibitem[Phung \latin{et~al.}(2020)Phung, Komori, Yanai, Sommerfeld, and
  Ehara]{phung20}
Phung,~Q.~M.; Komori,~Y.; Yanai,~T.; Sommerfeld,~T.; Ehara,~M. Combination of a
  {Voronoi}-Type Complex Absorbing Potential with the {XMS-CASPT2} Method and
  Pilot Applications. \emph{J. Chem. Theory Comput.} \textbf{2020}, \emph{16},
  2606--2616\relax
\mciteBstWouldAddEndPuncttrue
\mciteSetBstMidEndSepPunct{\mcitedefaultmidpunct}
{\mcitedefaultendpunct}{\mcitedefaultseppunct}\relax
\EndOfBibitem
\bibitem[Gayvert and Bravaya(2022)Gayvert, and Bravaya]{gayvert22a}
Gayvert,~J.~R.; Bravaya,~K.~B. Application of Box and {Voronoi CAPs} for
  Metastable Electronic States in Molecular Clusters. \emph{J. Phys. Chem. A}
  \textbf{2022}, \emph{126}, 5070--5078\relax
\mciteBstWouldAddEndPuncttrue
\mciteSetBstMidEndSepPunct{\mcitedefaultmidpunct}
{\mcitedefaultendpunct}{\mcitedefaultseppunct}\relax
\EndOfBibitem
\bibitem[Gayvert and Bravaya(2022)Gayvert, and Bravaya]{gayvert22b}
Gayvert,~J.~R.; Bravaya,~K.~B. Projected {CAP-EOM-CCSD} method for electronic
  resonances. \emph{J. Chem. Phys.} \textbf{2022}, \emph{156}, 094108\relax
\mciteBstWouldAddEndPuncttrue
\mciteSetBstMidEndSepPunct{\mcitedefaultmidpunct}
{\mcitedefaultendpunct}{\mcitedefaultseppunct}\relax
\EndOfBibitem
\bibitem[Dempwolff \latin{et~al.}(2021)Dempwolff, Belogolova, Sommerfeld,
  Trofimov, and Dreuw]{dempwolff22}
Dempwolff,~A.~L.; Belogolova,~A.~M.; Sommerfeld,~T.; Trofimov,~A.~B.; Dreuw,~A.
  {CAP/EA-ADC} method for metastable anions: Computational aspects and
  application to $\pi^*$ resonances of norbornadiene and 1,4-cyclohexadiene.
  \emph{J. Chem. Phys.} \textbf{2021}, \emph{155}, 054103\relax
\mciteBstWouldAddEndPuncttrue
\mciteSetBstMidEndSepPunct{\mcitedefaultmidpunct}
{\mcitedefaultendpunct}{\mcitedefaultseppunct}\relax
\EndOfBibitem
\bibitem[Stanton and Bartlett(1993)Stanton, and Bartlett]{stanton93}
Stanton,~J.~F.; Bartlett,~R.~J. The equation of motion coupled-cluster method.
  A systematic biorthogonal approach to molecular excitation energies,
  transition probabilities, and excited state properties. \emph{J. Chem. Phys.}
  \textbf{1993}, \emph{98}, 7029--7039\relax
\mciteBstWouldAddEndPuncttrue
\mciteSetBstMidEndSepPunct{\mcitedefaultmidpunct}
{\mcitedefaultendpunct}{\mcitedefaultseppunct}\relax
\EndOfBibitem
\bibitem[Nooijen and Bartlett(1996)Nooijen, and Bartlett]{nooijen96}
Nooijen,~M.; Bartlett,~R.~J. Equation of motion coupled cluster method for
  electron attachment. \emph{J. Chem. Phys.} \textbf{1996}, \emph{102},
  3629--3647\relax
\mciteBstWouldAddEndPuncttrue
\mciteSetBstMidEndSepPunct{\mcitedefaultmidpunct}
{\mcitedefaultendpunct}{\mcitedefaultseppunct}\relax
\EndOfBibitem
\bibitem[Epifanovsky \latin{et~al.}(2021)Epifanovsky, Gilbert, Feng, Lee, Mao,
  Mardirossian, Pokhilko, White, Coons, Dempwolff, Gan, Hait, Horn, Jacobson,
  Kaliman, Kussmann, Lange, Lao, Levine, Liu, McKenzie, Morrison, Nanda,
  Plasser, Rehn, Vidal, You, Zhu, Alam, Albrecht, Aldossary, Alguire, Andersen,
  Athavale, Barton, Begam, Behn, Bellonzi, Bernard, Berquist, Burton, Carreras,
  Carter-Fenk, Chakraborty, Chien, Closser, Cofer-Shabica, Dasgupta,
  de~Wergifosse, Deng, Diedenhofen, Do, Ehlert, Fang, Fatehi, Feng, Friedhoff,
  Gayvert, Ge, Gidofalvi, Goldey, Gomes, Gonz\'{a}lez-Espinoza, Gulania,
  Gunina, Hanson-Heine, Harbach, Hauser, Herbst, Hern\'{a}ndez~Vera, Hodecker,
  Holden, Houck, Huang, Hui, Huynh, Ivanov, J\'{a}sz, Ji, Jiang, Kaduk,
  Kähler, Khistyaev, Kim, Kis, Klunzinger, Koczor-Benda, Koh, Kosenkov,
  Koulias, Kowalczyk, Krauter, Kue, Kunitsa, Kus, Ladj\'{a}nszki, Landau,
  Lawler, Lefrancois, Lehtola, Li, Li, Liang, Liebenthal, Lin, Lin, Liu, Liu,
  Loipersberger, Luenser, Manjanath, Manohar, Mansoor, Manzer, Mao, Marenich,
  Markovich, Mason, Maurer, McLaughlin, Menger, Mewes, Mewes, Morgante,
  Mullinax, Oosterbaan, Paran, Paul, Paul, Pavo\v{s}evi\'{c}, Pei, Prager,
  Proynov, R\'{a}k, Ramos-Cordoba, Rana, Rask, Rettig, Richard, Rob, Rossomme,
  Scheele, Scheurer, Schneider, Sergueev, Sharada, Skomorowski, Small, Stein,
  Su, Sundstrom, Tao, Thirman, Tornai, Tsuchimochi, Tubman, Veccham, Vydrov,
  Wenzel, Witte, Yamada, Yao, Yeganeh, Yost, Zech, Zhang, Zhang, Zhang, Zuev,
  Aspuru-Guzik, Bell, Besley, Bravaya, Brooks, Casanova, Chai, Coriani, Cramer,
  Cserey, DePrince, DiStasio, Dreuw, Dunietz, Furlani, Goddard,
  Hammes-Schiffer, Head-Gordon, Hehre, Hsu, Jagau, Jung, Klamt, Kong,
  Lambrecht, Liang, Mayhall, McCurdy, Neaton, Ochsenfeld, Parkhill, Peverati,
  Rassolov, Shao, Slipchenko, Stauch, Steele, Subotnik, Thom, Tkatchenko,
  Truhlar, Van~Voorhis, Wesolowski, Whaley, Woodcock, Zimmerman, Faraji, Gill,
  Head-Gordon, Herbert, and Krylov]{qchem50}
Epifanovsky,~E.; Gilbert,~A. T.~B.; Feng,~X.; Lee,~J.; Mao,~Y.;
  Mardirossian,~N.; Pokhilko,~P.; White,~A.~F.; Coons,~M.~P.; Dempwolff,~A.~L.;
  Gan,~Z.; Hait,~D.; Horn,~P.~R.; Jacobson,~L.~D.; Kaliman,~I.; Kussmann,~J.;
  Lange,~A.~W.; Lao,~K.~U.; Levine,~D.~S.; Liu,~J.; McKenzie,~S.~C.;
  Morrison,~A.~F.; Nanda,~K.~D.; Plasser,~F.; Rehn,~D.~R.; Vidal,~M.~L.;
  You,~Z.-Q.; Zhu,~Y.; Alam,~B.; Albrecht,~B.~J.; Aldossary,~A.; Alguire,~E.;
  Andersen,~J.~H.; Athavale,~V.; Barton,~D.; Begam,~K.; Behn,~A.; Bellonzi,~N.;
  Bernard,~Y.~A.; Berquist,~E.~J.; Burton,~H. G.~A.; Carreras,~A.;
  Carter-Fenk,~K.; Chakraborty,~R.; Chien,~A.~D.; Closser,~K.~D.;
  Cofer-Shabica,~V.; Dasgupta,~S.; de~Wergifosse,~M.; Deng,~J.;
  Diedenhofen,~M.; Do,~H.; Ehlert,~S.; Fang,~P.-T.; Fatehi,~S.; Feng,~Q.;
  Friedhoff,~T.; Gayvert,~J.; Ge,~Q.; Gidofalvi,~G.; Goldey,~M.; Gomes,~J.;
  Gonz\'{a}lez-Espinoza,~C.~E.; Gulania,~S.; Gunina,~A.~O.; Hanson-Heine,~M.
  W.~D.; Harbach,~P. H.~P.; Hauser,~A.; Herbst,~M.~F.; Hern\'{a}ndez~Vera,~M.;
  Hodecker,~M.; Holden,~Z.~C.; Houck,~S.; Huang,~X.; Hui,~K.; Huynh,~B.~C.;
  Ivanov,~M.; J\'{a}sz,~A.; Ji,~H.; Jiang,~H.; Kaduk,~B.; Kähler,~S.;
  Khistyaev,~K.; Kim,~J.; Kis,~G.; Klunzinger,~P.; Koczor-Benda,~Z.;
  Koh,~J.~H.; Kosenkov,~D.; Koulias,~L.; Kowalczyk,~T.; Krauter,~C.~M.;
  Kue,~K.; Kunitsa,~A.; Kus,~T.; Ladj\'{a}nszki,~I.; Landau,~A.; Lawler,~K.~V.;
  Lefrancois,~D.; Lehtola,~S.; Li,~R.~R.; Li,~Y.-P.; Liang,~J.; Liebenthal,~M.;
  Lin,~H.-H.; Lin,~Y.-S.; Liu,~F.; Liu,~K.-Y.; Loipersberger,~M.; Luenser,~A.;
  Manjanath,~A.; Manohar,~P.; Mansoor,~E.; Manzer,~S.~F.; Mao,~S.-P.;
  Marenich,~A.~V.; Markovich,~T.; Mason,~S.; Maurer,~S.~A.; McLaughlin,~P.~F.;
  Menger,~M. F. S.~J.; Mewes,~J.-M.; Mewes,~S.~A.; Morgante,~P.;
  Mullinax,~J.~W.; Oosterbaan,~K.~J.; Paran,~G.; Paul,~A.~C.; Paul,~S.~K.;
  Pavo\v{s}evi\'{c},~F.; Pei,~Z.; Prager,~S.; Proynov,~E.~I.; R\'{a}k,~a.;
  Ramos-Cordoba,~E.; Rana,~B.; Rask,~A.~E.; Rettig,~A.; Richard,~R.~M.;
  Rob,~F.; Rossomme,~E.; Scheele,~T.; Scheurer,~M.; Schneider,~M.;
  Sergueev,~N.; Sharada,~S.~M.; Skomorowski,~W.; Small,~D.~W.; Stein,~C.~J.;
  Su,~Y.-C.; Sundstrom,~E.~J.; Tao,~Z.; Thirman,~J.; Tornai,~G.~J.;
  Tsuchimochi,~T.; Tubman,~N.~M.; Veccham,~S.~P.; Vydrov,~O.; Wenzel,~J.;
  Witte,~J.; Yamada,~A.; Yao,~K.; Yeganeh,~S.; Yost,~S.~R.; Zech,~A.;
  Zhang,~I.~Y.; Zhang,~X.; Zhang,~Y.; Zuev,~D.; Aspuru-Guzik,~A.; Bell,~A.~T.;
  Besley,~N.~A.; Bravaya,~K.~B.; Brooks,~B.~R.; Casanova,~D.; Chai,~J.-D.;
  Coriani,~S.; Cramer,~C.~J.; Cserey,~G.; DePrince,~A.~E.; DiStasio,~R.~A.;
  Dreuw,~A.; Dunietz,~B.~D.; Furlani,~T.~R.; Goddard,~W.~A.;
  Hammes-Schiffer,~S.; Head-Gordon,~T.; Hehre,~W.~J.; Hsu,~C.-P.; Jagau,~T.-C.;
  Jung,~Y.; Klamt,~A.; Kong,~J.; Lambrecht,~D.~S.; Liang,~W.; Mayhall,~N.~J.;
  McCurdy,~C.~W.; Neaton,~J.~B.; Ochsenfeld,~C.; Parkhill,~J.~A.; Peverati,~R.;
  Rassolov,~V.~A.; Shao,~Y.; Slipchenko,~L.~V.; Stauch,~T.; Steele,~R.~P.;
  Subotnik,~J.~E.; Thom,~A. J.~W.; Tkatchenko,~A.; Truhlar,~D.~G.;
  Van~Voorhis,~T.; Wesolowski,~T.~A.; Whaley,~K.~B.; Woodcock,~H.~L.;
  Zimmerman,~P.~M.; Faraji,~S.; Gill,~P. M.~W.; Head-Gordon,~M.;
  Herbert,~J.~M.; Krylov,~A.~I. Software for the frontiers of quantum
  chemistry: An overview of developments in the {Q-Chem} 5 package. \emph{J.
  Chem. Phys.} \textbf{2021}, \emph{155}, 084801\relax
\mciteBstWouldAddEndPuncttrue
\mciteSetBstMidEndSepPunct{\mcitedefaultmidpunct}
{\mcitedefaultendpunct}{\mcitedefaultseppunct}\relax
\EndOfBibitem
\bibitem[Gilbert \latin{et~al.}(2008)Gilbert, Besley, and Gill]{gilbert08}
Gilbert,~A. T.~B.; Besley,~N.~A.; Gill,~P. M.~W. Self-Consistent Field
  Calculations of Excited States Using the Maximum Overlap Method ({MOM}).
  \emph{J. Phys. Chem. A} \textbf{2008}, \emph{112}, 13164--13171\relax
\mciteBstWouldAddEndPuncttrue
\mciteSetBstMidEndSepPunct{\mcitedefaultmidpunct}
{\mcitedefaultendpunct}{\mcitedefaultseppunct}\relax
\EndOfBibitem
\bibitem[Becke(1988)]{becke88}
Becke,~A.~D. A multicenter numerical integration scheme for polyatomic
  molecules. \emph{J. Chem. Phys.} \textbf{1988}, \emph{88}, 2547--2553\relax
\mciteBstWouldAddEndPuncttrue
\mciteSetBstMidEndSepPunct{\mcitedefaultmidpunct}
{\mcitedefaultendpunct}{\mcitedefaultseppunct}\relax
\EndOfBibitem
\bibitem[Berman \latin{et~al.}(1983)Berman, Estrada, Cederbaum, and
  Domcke]{berman83}
Berman,~M.; Estrada,~H.; Cederbaum,~L.~S.; Domcke,~W. Nuclear dynamics in
  resonant electron-molecule scattering beyond the local approximation: The
  2.3-{eV} shape resonance in {N$_2$}. \emph{Phys. Rev. A} \textbf{1983},
  \emph{28}, 1363--1381\relax
\mciteBstWouldAddEndPuncttrue
\mciteSetBstMidEndSepPunct{\mcitedefaultmidpunct}
{\mcitedefaultendpunct}{\mcitedefaultseppunct}\relax
\EndOfBibitem
\bibitem[Jagau and Krylov(2016)Jagau, and Krylov]{jagau16}
Jagau,~T.-C.; Krylov,~A.~I. Characterizing metastable states beyond energies
  and lifetimes: Dyson orbitals and transition dipole moments. \emph{J. Chem.
  Phys.} \textbf{2016}, \emph{144}, 054113\relax
\mciteBstWouldAddEndPuncttrue
\mciteSetBstMidEndSepPunct{\mcitedefaultmidpunct}
{\mcitedefaultendpunct}{\mcitedefaultseppunct}\relax
\EndOfBibitem
\bibitem[Walker \latin{et~al.}(1978)Walker, Stamatovic, and Wong]{walker78}
Walker,~I.~C.; Stamatovic,~A.; Wong,~S.~F. Vibrational excitation of ethylene
  by electron impact: 1-11 {eV}. \emph{J. Chem. Phys.} \textbf{1978},
  \emph{69}, 5532--5537\relax
\mciteBstWouldAddEndPuncttrue
\mciteSetBstMidEndSepPunct{\mcitedefaultmidpunct}
{\mcitedefaultendpunct}{\mcitedefaultseppunct}\relax
\EndOfBibitem
\bibitem[Panajotovic \latin{et~al.}(2003)Panajotovic, Kitajima, Tanaka,
  Jelisavcic, Lower, Campbell, Brunger, and Buckman]{panajotovic03}
Panajotovic,~R.; Kitajima,~M.; Tanaka,~H.; Jelisavcic,~M.; Lower,~J.;
  Campbell,~L.; Brunger,~M.~J.; Buckman,~S.~J. Electron collisions with
  ethylene. \emph{J. Phys. B: At. Mol. Opt. Phys.} \textbf{2003}, \emph{36},
  1615--1626\relax
\mciteBstWouldAddEndPuncttrue
\mciteSetBstMidEndSepPunct{\mcitedefaultmidpunct}
{\mcitedefaultendpunct}{\mcitedefaultseppunct}\relax
\EndOfBibitem
\bibitem[White \latin{et~al.}(2017)White, Epifanovsky, McCurdy, and
  Head-Gordon]{white17}
White,~A.~F.; Epifanovsky,~E.; McCurdy,~C.~W.; Head-Gordon,~M. Second order
  {M{\o}ller-Plesset} and coupled cluster singles and doubles methods with
  complex basis functions for resonances in electron-molecule scattering.
  \emph{J. Chem. Phys.} \textbf{2017}, \emph{146}, 234107\relax
\mciteBstWouldAddEndPuncttrue
\mciteSetBstMidEndSepPunct{\mcitedefaultmidpunct}
{\mcitedefaultendpunct}{\mcitedefaultseppunct}\relax
\EndOfBibitem
\bibitem[Aflatooni \latin{et~al.}(2001)Aflatooni, Hitt, Gallup, and
  Burrow]{aflatooni01}
Aflatooni,~K.; Hitt,~B.; Gallup,~G.~A.; Burrow,~P.~D. Temporary anion states of
  selected amino acids. \emph{J. Chem. Phys.} \textbf{2001}, \emph{115},
  6489--6494\relax
\mciteBstWouldAddEndPuncttrue
\mciteSetBstMidEndSepPunct{\mcitedefaultmidpunct}
{\mcitedefaultendpunct}{\mcitedefaultseppunct}\relax
\EndOfBibitem
\bibitem[Hocking(1976)]{hocking76}
Hocking,~W.~H. The Other Rotamer of Formic Acid, cis-HCOOH. \emph{Z.
  Naturforsch. A} \textbf{1976}, \emph{31}, 1113--1121\relax
\mciteBstWouldAddEndPuncttrue
\mciteSetBstMidEndSepPunct{\mcitedefaultmidpunct}
{\mcitedefaultendpunct}{\mcitedefaultseppunct}\relax
\EndOfBibitem
\bibitem[Rescigno \latin{et~al.}(2006)Rescigno, Trevisan, and Orel]{rescigno06}
Rescigno,~T.~N.; Trevisan,~C.~S.; Orel,~A.~E. Dynamics of Low-Energy Electron
  Attachment to Formic Acid. \emph{Phys. Rev. Lett.} \textbf{2006}, \emph{96},
  213201\relax
\mciteBstWouldAddEndPuncttrue
\mciteSetBstMidEndSepPunct{\mcitedefaultmidpunct}
{\mcitedefaultendpunct}{\mcitedefaultseppunct}\relax
\EndOfBibitem
\bibitem[Pelc \latin{et~al.}(2002)Pelc, Sailer, Scheier, Probst, Mason,
  Illenberger, and M\"ark]{pelc02}
Pelc,~A.; Sailer,~W.; Scheier,~P.; Probst,~M.; Mason,~N.~J.; Illenberger,~E.;
  M\"ark,~T.~D. Dissociative electron attachment to formic acid ({HCOOH}).
  \emph{Chem. Phys. Lett.} \textbf{2002}, \emph{361}, 277--284\relax
\mciteBstWouldAddEndPuncttrue
\mciteSetBstMidEndSepPunct{\mcitedefaultmidpunct}
{\mcitedefaultendpunct}{\mcitedefaultseppunct}\relax
\EndOfBibitem
\bibitem[Allan(2006)]{allan06}
Allan,~M. Study of resonances in formic acid by means of vibrational excitation
  by slow electrons. \emph{J. Phys. B: At. Mol. Opt. Phys.} \textbf{2006},
  \emph{39}, 2939--2947\relax
\mciteBstWouldAddEndPuncttrue
\mciteSetBstMidEndSepPunct{\mcitedefaultmidpunct}
{\mcitedefaultendpunct}{\mcitedefaultseppunct}\relax
\EndOfBibitem
\bibitem[Vizcaino \latin{et~al.}(2006)Vizcaino, Jelisavcic, Sullivan, and
  Buckman]{vizcaino06}
Vizcaino,~V.; Jelisavcic,~M.; Sullivan,~J.~P.; Buckman,~S.~J. Elastic electron
  scattering from formic acid ({HCOOH}): absolute differential cross-sections.
  \emph{New J. Phys.} \textbf{2006}, \emph{8}, 85\relax
\mciteBstWouldAddEndPuncttrue
\mciteSetBstMidEndSepPunct{\mcitedefaultmidpunct}
{\mcitedefaultendpunct}{\mcitedefaultseppunct}\relax
\EndOfBibitem
\bibitem[Dunning(1989)]{dunning89}
Dunning,~T.~H. Gaussian basis sets for use in correlated molecular
  calculations. {I}. The atoms boron through neon and hydrogen. \emph{J. Chem.
  Phys.} \textbf{1989}, \emph{90}, 1007--1023\relax
\mciteBstWouldAddEndPuncttrue
\mciteSetBstMidEndSepPunct{\mcitedefaultmidpunct}
{\mcitedefaultendpunct}{\mcitedefaultseppunct}\relax
\EndOfBibitem
\bibitem[Kendall \latin{et~al.}(1992)Kendall, Dunning, and Harrison]{kendall92}
Kendall,~R.~A.; Dunning,~T.~H.; Harrison,~R.~J. Electron affinities of the
  first‐row atoms revisited. Systematic basis sets and wave functions.
  \emph{J. Chem. Phys.} \textbf{1992}, \emph{96}, 6796--6806\relax
\mciteBstWouldAddEndPuncttrue
\mciteSetBstMidEndSepPunct{\mcitedefaultmidpunct}
{\mcitedefaultendpunct}{\mcitedefaultseppunct}\relax
\EndOfBibitem
\end{mcitethebibliography}

\end{document}